\documentclass[11pt]{article} 
\usepackage[utf8]{inputenc} 
\usepackage[a4paper, margin=1in]{geometry} 
\usepackage{titlesec} 
\titleformat{\paragraph}[hang]{\normalfont\bfseries\itshape}{}{0pt}{}
\titlespacing*{\paragraph}{0pt}{2.25ex plus 1ex minus .2ex}{1ex plus .2ex}
\usepackage{graphicx} 
\usepackage{caption}  
\usepackage{fancyhdr} 
\usepackage{amsmath, amssymb} 
\usepackage{booktabs} 
\usepackage{longtable} 
\usepackage{float} 
\usepackage{mathptmx} 
\usepackage{array} 
\usepackage[hidelinks]{hyperref}

\usepackage{listings}
\graphicspath{{results_and_discussion/u238_doppler_vs_openmc/}%
              {results_and_discussion/benchmarks_and_data_prep/figures/}%
              {results_and_discussion/ablation/}}

\title{Outram Park Part II: Agentic Verification, Validation and Documentation as a Human Review Gate for Nuclear Data and Neutronics Software}

\author{Theodore Kay Chen Ong, SiCong Xiao}
\date{19 October 2026}

\begin{document}

\maketitle

\begin{abstract}
NJOY21 \cite{njoy21} is part of a modernisation effort for NJOY2016, but consists mostly of C++ wrappers. Agentic coding represents a unique opportunity for NJOY modernisation in Rust, but agents require human expert supervision, which may be in short supply. Moreover, expert human review is the bottleneck in agentic software development as human resources are limited in this context. This work describes a methodology for how a human manager of AI agents can reasonably translate and refactor scientific software with limited domain expertise using principles borrowed from nuclear quality assurance documents. Moreover, it presents a methodology for when to allocate limited human resources for reviewing the agentically generated code. This methodology centres on agentic V\&V and documentation, and it is demonstrated via construction of Monte Carlo web demos for the Rust fork of NJOY and Outram Monte Carlo (OutramMC), Rust ports of NJOY2016 and OpenMC, in Outram Park. In this work, an integrated pipeline for the Rust port of NJOY and OutramMC was demonstrated in web demos, and code-to-code verification and validation was performed with ICSBEP benchmarks for Godiva, HST009 and LCT008. ACE files produced with Outram Park's NJOY port were identical for more than 30 ACE files. These web demos also served as part of agentic documentation, which prepared the codebase for human review.
\end{abstract}

\section{Introduction}

Artificial intelligence (AI) large language models (LLMs) and coding agents are increasingly capable of generating, translating, debugging and testing software. These capabilities provide opportunities for accelerating scientific software development, particularly where established scientific codes already exist and can serve as upstream references or numerical oracles. Rather than requiring a human developer to implement every numerical method manually, AI agents can be instructed to translate existing implementations, construct tests against upstream software, identify discrepancies and iteratively debug the resulting code.

Scientific software, however, presents additional challenges compared with general software development \cite{kanewala2014testing,peng2021unit}. A program that compiles and executes successfully does not necessarily implement the correct scientific model. Likewise, numerical results that appear reasonable, or even agree with expected values, do not necessarily demonstrate that the underlying physical models and numerical methods have been implemented correctly. This problem becomes particularly important for agentically generated software because an AI agent can iteratively modify software until a requested numerical target is reached without necessarily preserving the intended physics.

Consequently, verification and validation (V\&V) become particularly important for agentically developed scientific software. Where reliable upstream implementations exist, these provide numerical oracles against which individual functions, integrated calculations and complete simulation workflows can be compared. Differential testing can therefore provide a systematic mechanism for detecting discrepancies between an agentically developed implementation and its upstream reference \cite{mckeeman1998differential}.

Nuclear engineering provides a useful domain in which to investigate these issues. Nuclear engineering software frequently involves coupled numerical methods, evaluated nuclear data, stochastic simulation and safety-relevant physical models. At the same time, mature open-source and publicly available scientific software provides established reference implementations against which newly developed software can be compared. Nuclear engineering also has established software quality assurance practices through standards and regulatory guidance such as ASME NQA-1 \cite{kendrick2019nuclear} and NUREG/BR-0167 \cite{nureg_br_0167}.

Outram Park is an open-source scientific software project intended to provide an integrated simulation environment in Rust. As part of its development, established scientific software has been translated or reimplemented with extensive assistance from AI coding agents. Two important components are a Rust port/fork of NJOY for nuclear data processing and OutramMC, a Rust implementation derived from OpenMC for Monte Carlo neutron transport. There is particular motivation to translate the former because NJOY2016 was considered outdated, and NJOY21 attempted to wrap it in C++ but was not a true translation to C++ to modernise it \cite{njoy21}. Translation of NJOY2016 algorithms to Rust would indeed be valuable as it has the potential to make the eventual fork of NJOY more user friendly, especially when it can be integrated as a Rust dependency for the Rust port of OpenMC (OutramMC). Together, these provide a useful case study spanning nuclear data processing, numerical algorithms, stochastic neutron transport and criticality benchmark calculations.

The translation of mature scientific software also provides an unusual opportunity for agentic software engineering. Unlike development of entirely new scientific algorithms, an upstream implementation already exists and can act as an oracle. Individual functions can be compared against upstream functions, processed nuclear data can be compared against established nuclear data processing software, and integrated neutron transport calculations can ultimately be compared against both upstream Monte Carlo codes and experimental criticality benchmarks.

\subsection{Literature Review}

Existing work on agentic software engineering demonstrates the increasing capability of AI agents to generate, translate, debug and modify substantial software systems. At the same time, the increasing autonomy of coding agents challenges traditional assumptions regarding human-readable code and exhaustive human code review. Ustynov discusses how conventional software-engineering practices may change in an era where software is increasingly produced and manipulated by agents \cite{ustynov2026beyond}, while Monperrus examines the possibility that coding agents may supersede traditional mandatory human code inspection \cite{monperrus2026codereview}. This raises an important question for scientific software: if humans can no longer reasonably inspect every line of agentically generated code, what evidence is required before such software should be subjected to and accepted through human review?

Human accountability nevertheless remains important when AI agents perform increasingly large portions of software development. In a speech to AI Engineer Singapore on 16 May 2026, Minister for Foreign Affairs Dr Vivian Balakrishnan emphasised that while calculations, computation, memory and other activities may increasingly be outsourced, personal understanding cannot be outsourced and accountability cannot be delegated \cite{balakrishnan2026aiengineer}. Thus, increased automation of software production does not necessarily imply automation of responsibility for the resulting scientific software.

Existing software-engineering research also suggests that meaningful human understanding does not necessarily require exhaustive inspection of every line of a large codebase. Huang and Madeira proposed targeted code inspection based on human errors, directing inspection effort toward portions of software more likely to warrant human attention \cite{huang2022targeted}. Baltes et al. similarly studied how developers navigate software when locating performance problems, demonstrating selective navigation and inspection of relevant portions of a codebase \cite{baltes2017navigate}. These approaches predate the current agentic coding era but provide useful precedent for selective human inspection of large agentically developed codebases.

Natural-language documentation may provide another intermediate layer between generated software and human understanding. Yuan et al. propose using ``unambiguous Natural Language (NL)-based documentation as a readable intermediate layer between prompts and code'' for human validation of LLM-generated software \cite{yuan2026verifiable}. Such an approach allows human reviewers to reason about intended behaviour before selectively examining the implementation responsible for that behaviour.

Interactive documentation provides another mechanism for making software understandable. Casdoc explored non-linear software documentation in which users interact with functional code examples rather than consuming documentation entirely as sequential text \cite{casdoc2025}. Interactive simulation also has precedent in physics education. Wieman et al. demonstrated the use of interactive simulations to support engagement and conceptual learning in physics \cite{wieman2008interactive}. These approaches suggest that scientific-software documentation need not consist solely of source-code documentation and textual manuals, but may also expose interactive demonstrations of the physical models implemented by the software. A notable relevant example is nucdatatools \cite{nucdatatools2026} wherein cross sections can be visualised in the browser. However, a full nuclear data processing to Monte Carlo pipeline was not yet done in nucdatatools, or deployed anywhere on the web as a web demo to the best of the author's knowledge. This will be attempted in this work.

Nuclear engineering provides a useful framework for addressing the verification problem because software quality assurance, verification and validation are already well established. Guidance such as NQA-1 \cite{kendrick2019nuclear} and NUREG/BR-0167 \cite{nureg_br_0167} separates requirements, implementation, unit and integration testing, qualification testing, acceptance testing and documentation. These practices provide a potential framework for adapting agentic software development to scientific applications without abandoning established expectations of scientific correctness and accountability.

There is also established precedent for differential verification of nuclear-data processing software. Tada et al. compared the FRENDY nuclear-data processing system against NJOY using processed nuclear data and downstream criticality calculations \cite{tada2017frendy}. Their work demonstrates that a newly developed nuclear-data processing implementation can be verified against an established processor both at the level of processed nuclear data and through its effects on integrated neutron transport calculations.

Monte Carlo criticality literature similarly provides established approaches for evaluating numerical agreement. Gao et al. used statistical agreement within two standard deviations when validating the RMC Monte Carlo code against the C5G7 benchmark and MCNP, providing precedent for a \(2\sigma\) comparison criterion; their eigenvalues agreed with MCNP to within 15 pcm \cite{gao2013rmc}. Criticality benchmark literature additionally provides evaluated experimental uncertainty bounds against which calculated \(k_\mathrm{eff}\) values can be compared. These provide quantitative references for distinguishing code-to-code verification from validation against experimental benchmarks.

In this context, agentic verification and validation need not replace human review. Instead, automated unit testing, integration testing, differential testing, qualification against established software and experimental benchmarks, and acceptance testing can progressively establish evidence that an agentically developed codebase is sufficiently mature to warrant human review. Human attention can then be directed toward scientifically important algorithms, assumptions, interfaces and anomalous results rather than requiring exhaustive inspection of every agent-generated line of code.

In this work, we explore the use of agentic unit testing, integration testing and differential testing as means of verifying and validating agentically developed scientific software where upstream oracles exist, even for cases where the human driver behind agents does not have high levels of domain expert knowledge. This makes the method applicable across scientific software so long as the upstream source code is available and produces reliable data. NJOY and OpenMC were selected as case studies because together they span nuclear data processing and Monte Carlo neutron transport, allowing the development process to be examined from individual numerical functions through integrated physics calculations and established criticality benchmarks. Moreover, the first author has little experience in developing these codes from the ground up; thus, it forms a good example wherein a human manager of AI agents can use these techniques to translate, test and verify code without extensive domain knowledge.

Through this case study, AI agents were used not only to translate and develop the software, but also to test, debug, document and deploy it. Once these steps are completed, it signals the code's readiness for human review, thus saving limited human resources and time on code which already functions as correctly as possible based on agentic development and testing. The methodology itself therefore forms the principal contribution of this work, while the resulting integrated Rust workflow from ENDF nuclear data processing through Monte Carlo transport and ICSBEP benchmark calculations serves as a practical demonstration of the approach.

\section{Methods}

\subsection{Case Study Selection and Motivation}

Open-source nuclear data processing and neutronics software are integral to Outram Park because they are required to calculate temperature feedback in reactors. Temperature feedback in turn underpins the safety case of a reactor during an anticipated transient without scram (ATWS). One postulated accident is the pressurised loss of forced cooling (PLOFC) in high-temperature gas-cooled reactors (HTGRs) without scram, in which the HTGR shuts down through its negative temperature feedback alone, without control rods \cite{gou2018thermal,kugeler2017htgrsafety}.

To calculate reactivity coefficients reliably in HTGRs, nuclear data must be Doppler broadened at multiple temperatures. This requires nuclear data processing software such as NJOY \cite{njoy2016} to be run across multiple temperatures. The same holds even for multigroup cross-section generation at fixed temperatures, where cross sections are interpolated between at least two temperatures. Moreover, Monte Carlo codes must be run to obtain k-eigenvalues at different fuel and moderator temperatures, and these codes should be able to perform full-core, doubly heterogeneous simulations. In OpenMC \cite{romano2015openmc}, this was not possible because OpenMC primarily uses surface tracking. While delta tracking and other methods had been submitted as pull requests \cite{wendt2026delta}, they had not been merged into mainline OpenMC as of September 2026.

OpenMC currently integrates with NJOY through a Python wrapper: NJOY converts ENDF files to compact ENDF (ACE) files, which are then converted to HDF5, the format OpenMC reads natively. The Python API abstracts these complex processes behind a simple interface. However, in the authors' experience, it neither offers the flexibility to modify the underlying NJOY algorithms nor exposes the full functionality of NJOY, because the API covers only a subset of NJOY's modules.

For instance, during the first author's PhD dissertation studies, he wanted to generate thermal scattering law data files from LEAPR input via the OpenMC Python API. This was not possible, as no such API existed. Thermal scattering laws are important for pebble-bed reactors such as fluoride-salt-cooled high-temperature reactors (FHRs) and HTGRs. This shows that while an API in another language can abstract away the complexity of the underlying code (NJOY), it must be constantly updated to expose its full functionality. In contrast, keeping the API within a single language removes this abstraction layer altogether. Even so, integration between Fortran and C++ would still be a challenge. It was therefore judged advantageous to translate both codebases to Rust and integrate them.

This provides strong motivation to translate NJOY and OpenMC to Rust, which offers modern language features, improved ease of use and memory safety compared with both C++ and Fortran \cite{nsa2023memorysafety,nsacisa2025memorysafelanguages}. Given that the first author has no experience in writing these codes from the ground up, this serves as an ideal case for testing the methodology described above.

\subsection{Requirements}

Given this context, it is important to define the goals of this endeavour; this is the first step of the software development process in NQA-1 \cite{kendrick2019nuclear}. In Outram Park, the goal is an integrated, cross-platform simulation system in one language that makes the workflow of deriving real-time surrogate models from high-fidelity models simple and smooth. As part of this goal, high-fidelity models with coupled multiphysics ought to be easy to construct and physically correct. This philosophy should be applied to advanced reactors of interest such as HTGRs.

Among HTGRs, it is easiest to start with test reactors for which abundant data are available in the open literature for code-to-code verification and validation. Test reactors such as the HTR-10 and HTTR were therefore suitable candidates \cite{iaea2003tecdoc1382}. As the HTR-10 had abundant geometry and material data available \cite{wu2002htr10}, it was chosen as the first demonstration case.

Among the multiphysics aspects, neutronics is a natural place to start, as the double heterogeneity problem is computationally challenging. Simulating it requires either stochastic geometry methods, such as the chord length sampling used in RMC \cite{liang2013chord,tan2025semi}, or delta tracking, as used in Serpent 2 \cite{leppanen2017use}. The prerequisite for this is the ability to perform nuclear data processing and basic Monte Carlo transport, as provided by the open-source codes NJOY and OpenMC. Hence, the requirement is that the translated codebase must fulfil the roles of NJOY and OpenMC, as far as neutron transport is concerned, in order to obtain benchmarked k-eigenvalue calculations. These must cover Godiva, the ``hello world'' of neutronics. They must also cover thermal-spectrum calculations, where thermal scattering and moderation through the resonance region become important. Cases were therefore chosen to span highly and low-enriched uranium, fast and thermal spectra, and homogeneous and heterogeneous fuel and moderator arrangements.

The following four ICSBEP cases \cite{briggs2003international} were chosen as the first targets for code-to-code verification and validation:

\begin{enumerate}
  \item Godiva (HEU-MET-FAST-001)
  \item Jemima (IEU-MET-FAST-002)
  \item HST-009 case 1
  \item LCT-008 case 1, initially as a simplified 11-nuclide model; since 14 October 2026 the full 35-nuclide model is used on every route
\end{enumerate}

\begin{figure}[H]
\centering
\includegraphics[width=\linewidth,height=0.75\textheight,keepaspectratio]{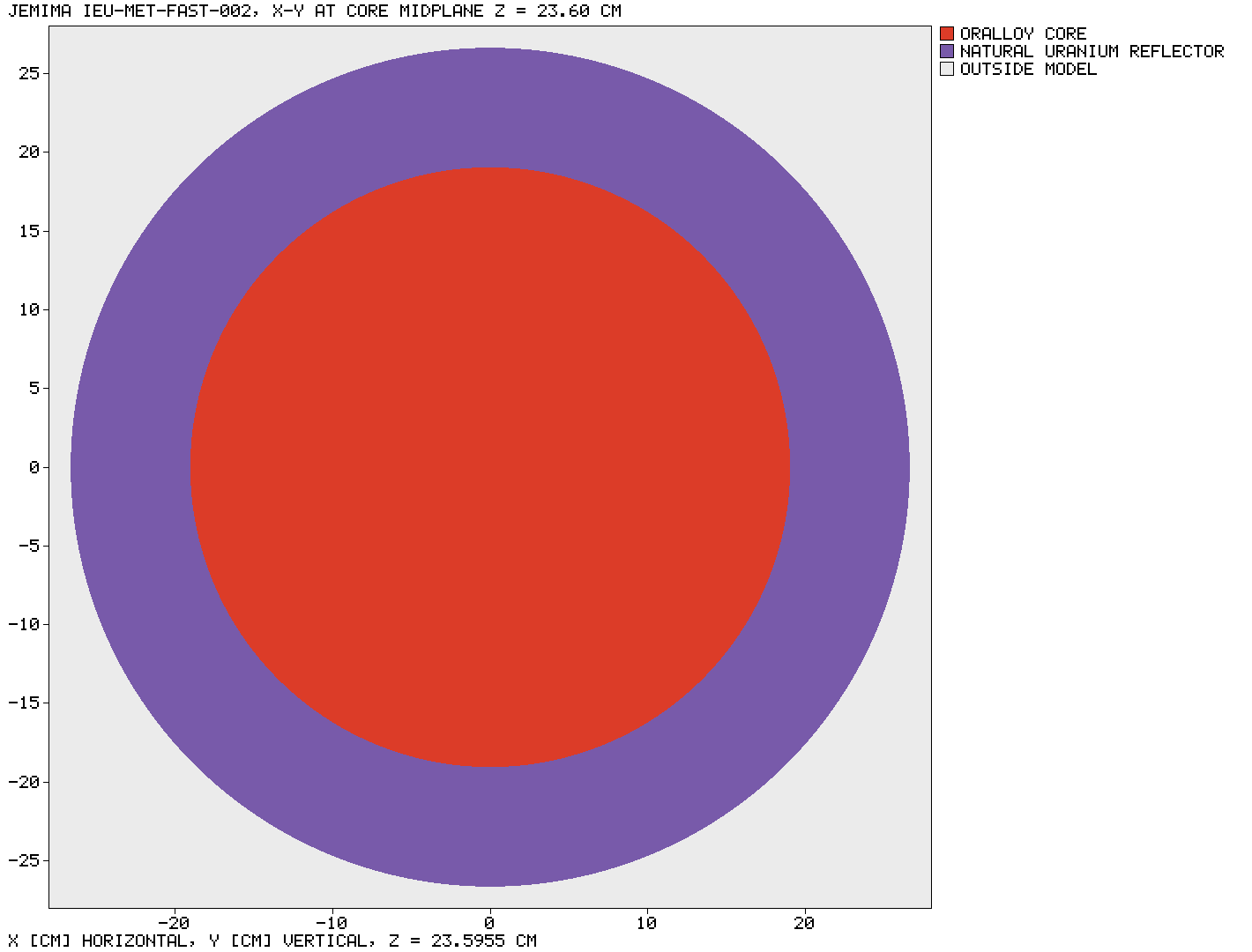}
\caption{Jemima (IEU-MET-FAST-002): x-y slice at the core midplane, z = 23.60 cm, showing the Oralloy core inside the natural uranium reflector.}
\label{fig:jemima_xy}
\end{figure}

\begin{figure}[H]
\centering
\includegraphics[width=\linewidth,height=0.75\textheight,keepaspectratio]{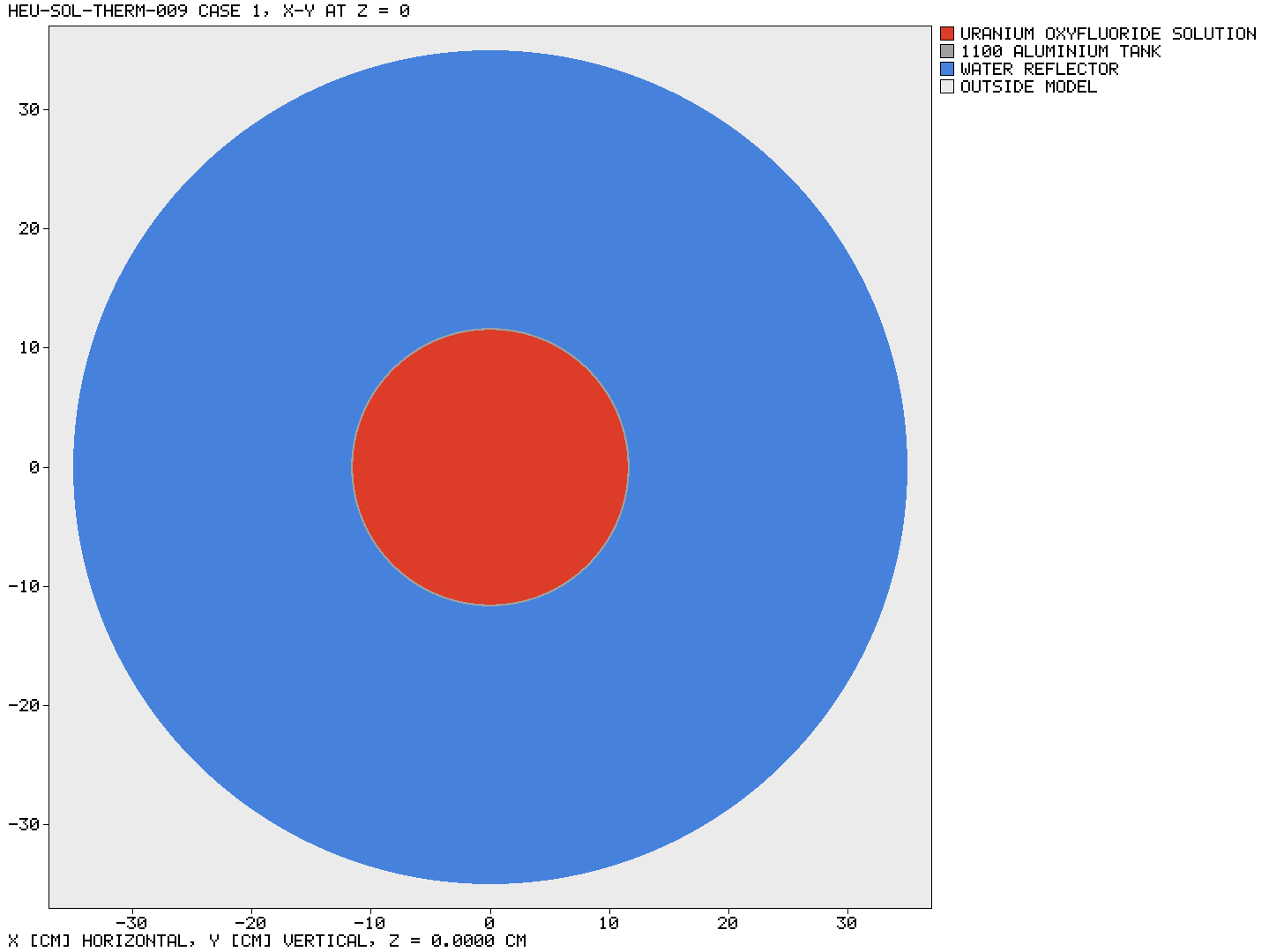}
\caption{HST-009 case 1 (HEU-SOL-THERM-009): x-y slice at z = 0, showing the uranium oxyfluoride solution in the 1100 aluminium tank, surrounded by the water reflector.}
\label{fig:hst009_xy}
\end{figure}

\begin{figure}[H]
\centering
\includegraphics[width=\linewidth,height=0.75\textheight,keepaspectratio]{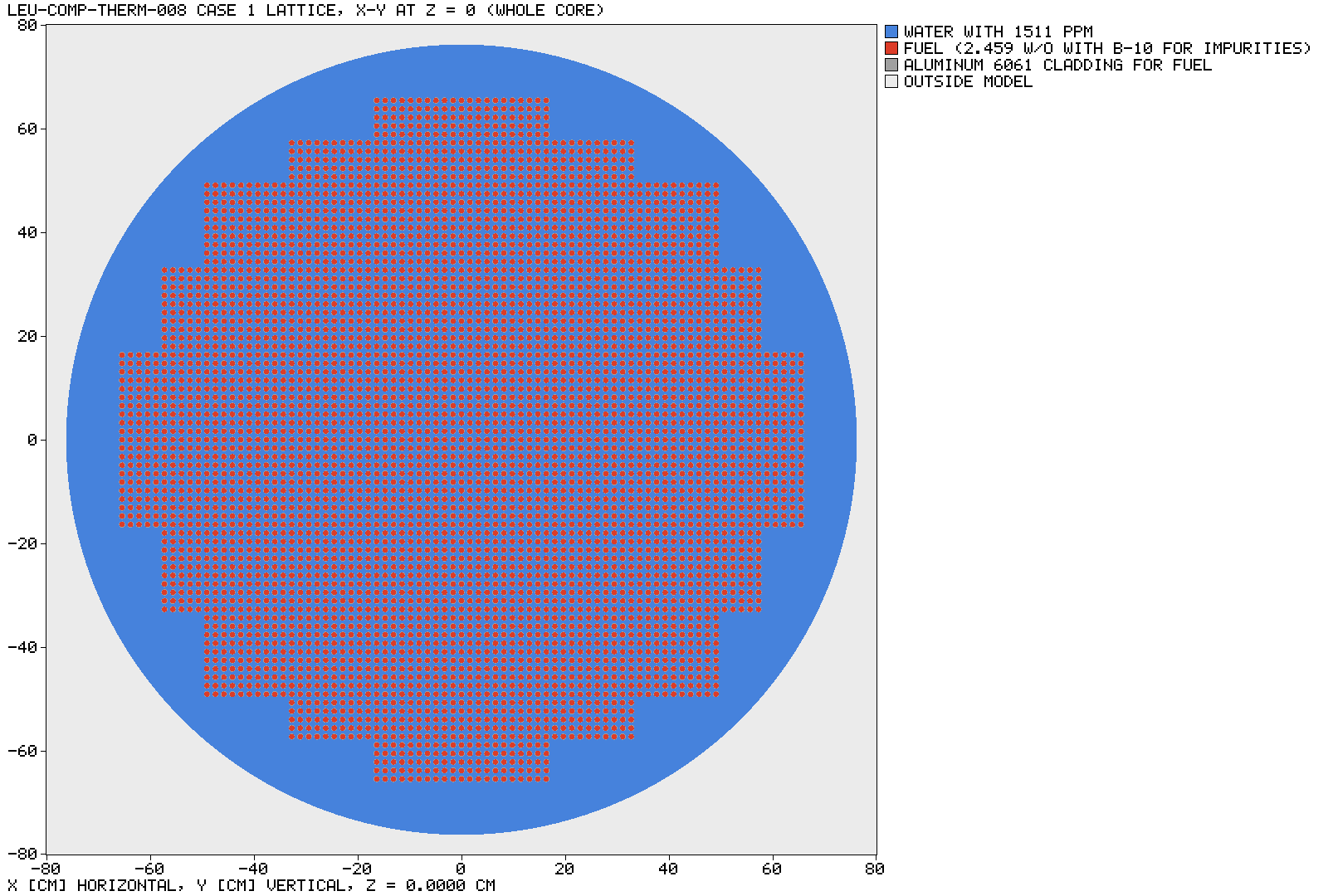}
\caption{LCT-008 case 1 (LEU-COMP-THERM-008): x-y slice at z = 0 of the whole core, showing the lattice of fuel rods (2.459 w/o, Al-6061 cladding) in water with 1511 ppm.}
\label{fig:lct008_xy_core}
\end{figure}

These cases do not require new delta-tracking algorithms; the surface-tracking algorithms in OpenMC suffice. Therefore, no new physics needs to be synthesised. Cross-sectional figures of these tests are shown in Figures \ref{fig:godiva_xy}, \ref{fig:jemima_xy}, \ref{fig:hst009_xy} and \ref{fig:lct008_xy_core}:

\begin{figure}[H]
\centering
\includegraphics[width=\linewidth,height=0.75\textheight,keepaspectratio]{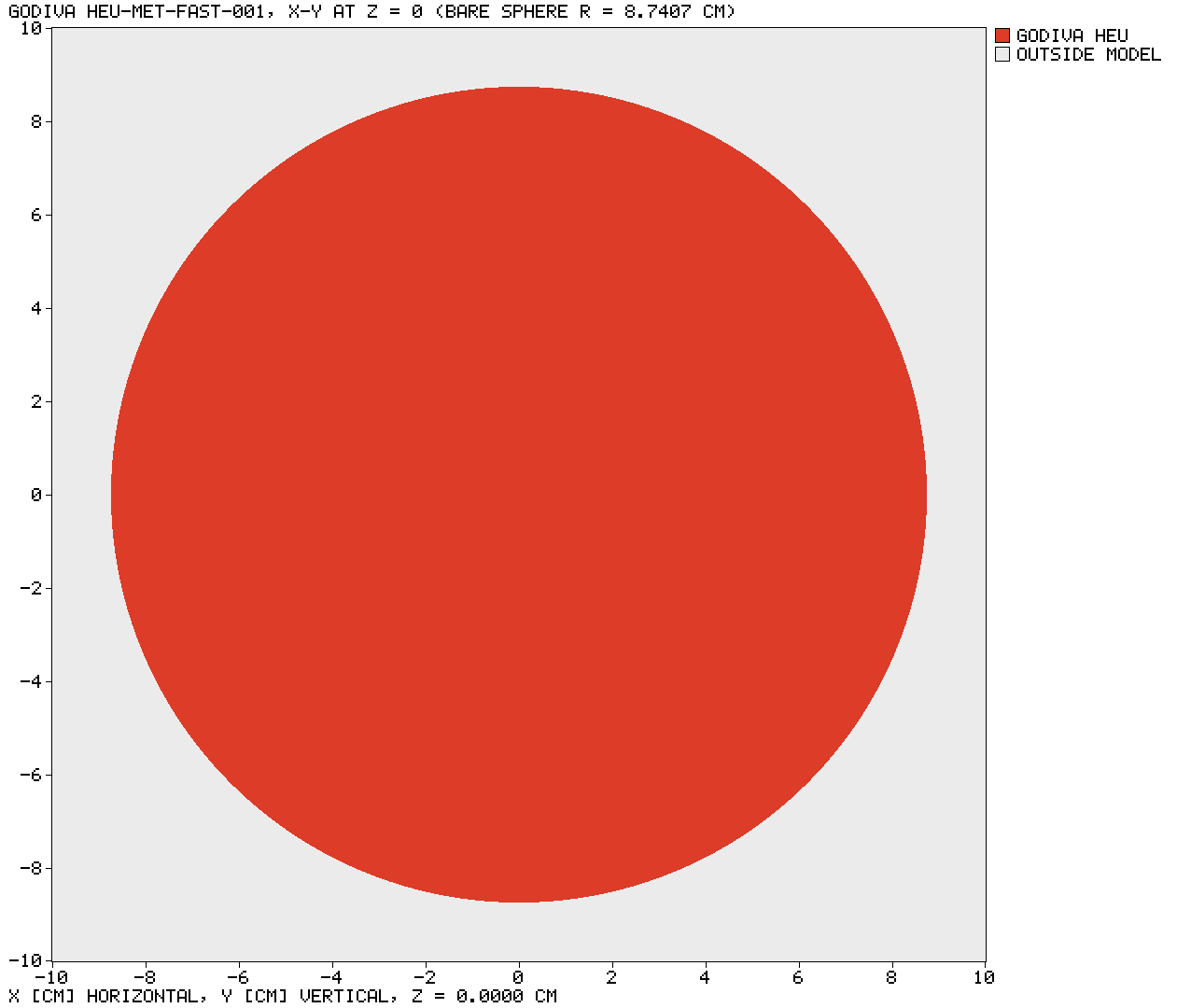}
\caption{Godiva (HEU-MET-FAST-001): x-y slice at z = 0 of the bare HEU sphere, R = 8.7407 cm, as modelled in this work.}
\label{fig:godiva_xy}
\end{figure}

These figures were generated using Rust ports of OpenMC's plotters, which demonstrates their capability of producing human-readable plots. For these four cases, results from NJOY and OpenMC are compared against the Rust port of NJOY and OutramMC (Outram Park's fork of OpenMC). To establish the Rust port's parity with NJOY, ACE files generated by the port are compared with those from NJOY2016. In addition, OpenMC runs using ACE files generated by the Rust port and by NJOY2016 should show statistically negligible differences. The evaluated benchmark values and uncertainties used in this work are:

\begingroup\small
\setlength{\tabcolsep}{3pt}
\begin{longtable}{>{\raggedright\arraybackslash}p{0.233\linewidth}>{\raggedleft\arraybackslash}p{0.233\linewidth}>{\raggedleft\arraybackslash}p{0.233\linewidth}>{\raggedright\arraybackslash}p{0.233\linewidth}}
\toprule
\textbf{Benchmark} & \textbf{Evaluated \(k_\mathrm{eff}\)} & \textbf{Uncertainty} & \textbf{Source} \\
\midrule
\endhead
Godiva (HEU-MET-FAST-001) & \(1.0000 \pm 0.0010\) & $\pm$100 pcm & Mosteller, Table 16-a \cite{mosteller2010expanded}; McSpaden et al. (2025) \cite{mcspaden2025music} \\
Jemima (IEU-MET-FAST-002) & \(1.0000 \pm 0.0030\) & $\pm$300 pcm & Mosteller, Table 17 \cite{mosteller2010expanded}; van der Marck, Table VI \cite{vandermarck2006benchmarking} \\
LCT-008 case 1 & \(1.0007 \pm 0.0016\) & $\pm$160 pcm & Mosteller, Table 18 \cite{mosteller2010expanded} \\
HST-009 case 1 & \(0.99900 \pm 0.00430\) & $\pm$430 pcm & van der Marck, Table IV \cite{vandermarck2006benchmarking} \\
\bottomrule
\end{longtable}
\endgroup

As a rule of thumb, we use the benchmark's uncertainty bounds and require the code-to-code computation to fall within those bounds. For code-to-code verification, the k-eigenvalues themselves have a distribution. Where two standard deviations of that distribution are smaller than the benchmark uncertainty bound, we use \(2\sigma\) as the passing criterion. Here, \(\sigma\) is the reported Monte Carlo standard deviation. However, we do not sample indefinitely, because Monte Carlo calculations are costly. As a rule of thumb, we therefore do not accept standard deviations above 100 pcm, even for cases with large experimental uncertainty.

\subsection{Agentic Porting and Testing Methodology}

The agentic porting methodology follows our previous work \cite{ong2026agenticportingconstructioninitial}, in which the upstream sources were cloned with git onto a local machine and AI agents were directed to translate the code to Rust. Testing and debugging followed. First, the agents were told to replicate the Godiva case, first with isotropic scattering and then with anisotropic scattering. Agents were also given pregenerated HDF5 files for U-238 and told to compare the NJOY port's output against them. This was the first level of verification: differential testing against NJOY-generated HDF5 files, together with the Godiva case. While these initial results were promising, they lacked the rigour required for proper V\&V of AI-generated software.

In early development, AI agents were instructed to construct a delta-tracked, HTR-10-based pebble bed, and were then asked to check which parameters had been coded wrongly so as to get as close as possible to the correct answers. This was close in concept to ablation testing. The agents constructed the pebble bed and the delta-tracking algorithms, which produced plausible-looking results. AI can produce plausible-looking results while getting the physics wrong. This brute-force testing (not quite ablation testing) yielded correct-looking results built on the wrong physics.

\begin{figure}[H]
\centering
\includegraphics[width=\linewidth,height=0.75\textheight,keepaspectratio]{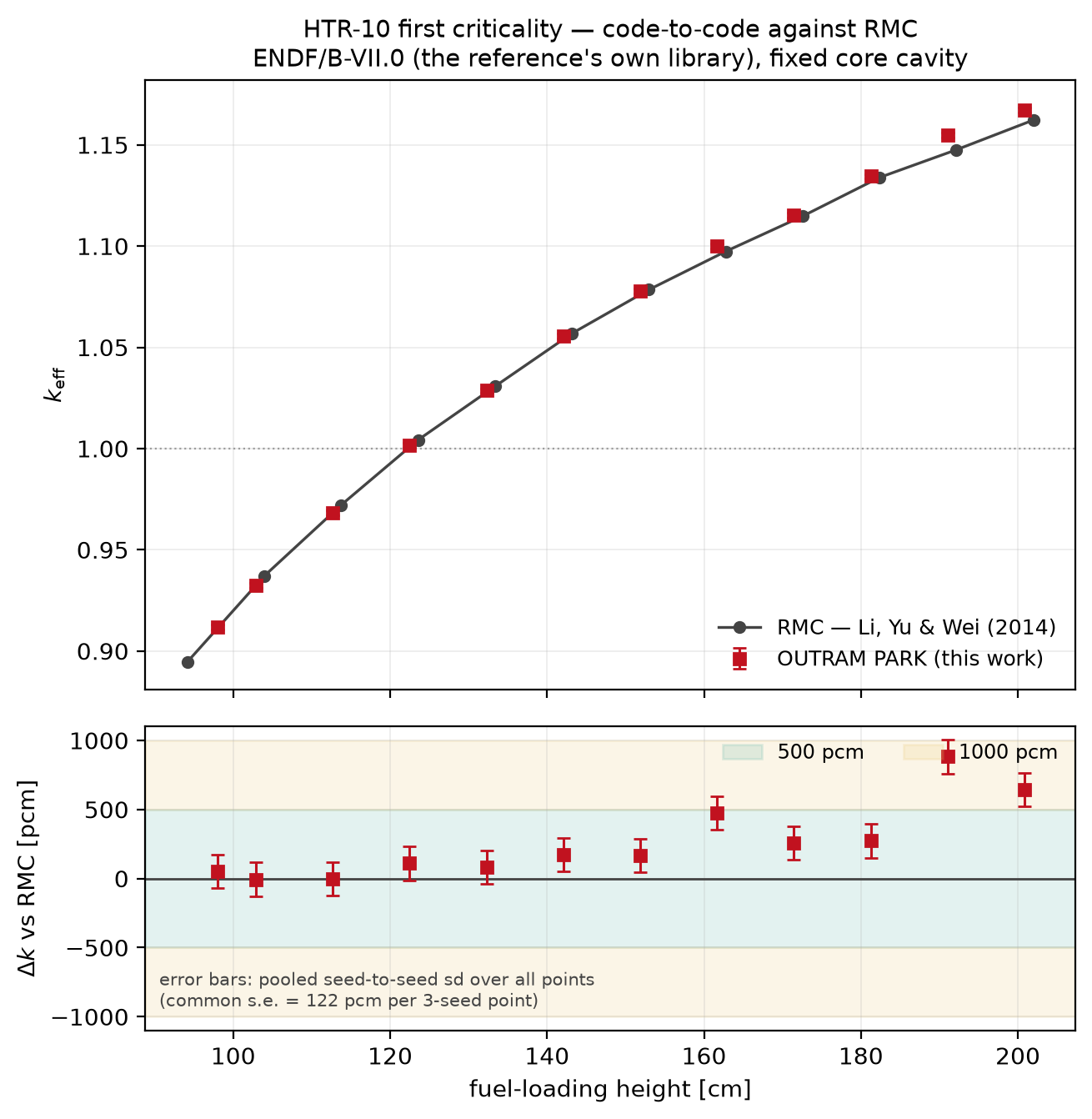}
\caption{Initial HTR-10 code-to-code comparison against RMC (Li, Yu and Wei, 2014): k\_eff against fuel-loading height, ENDF/B-VII.0, 3 seeds per height, on the first pebble-bed model (19 September 2026), before its interpenetrating pebbles were found. Every point lies within 1000 pcm of RMC. Data visualisation produced by a plotting script written with AI assistance (Claude Code) from the run data.}
\label{fig:htr10_rmc_curve}
\end{figure}

For example, in Figure \ref{fig:htr10_rmc_curve}, an HTR-10 bed was constructed in a code-to-code verification exercise against RMC, using the results of Li et al. \cite{li2014research}. In those results, RMC and MCNP differ by up to about 950 pcm (relative difference, Table 3 of \cite{li2014research}; 1089 pcm in absolute \(\Delta k\)) on this HTR-10 pebble-bed geometry. Since the OutramMC results were within 1000 pcm of RMC, the comparison looked plausibly correct. However, the underlying physics was wrong, because the geometry was wrong.

A subsequent geometry audit showed that the implemented HTR-10 pebble bed had dropped the A--B layer offset, so that neighbouring pebbles sat 4.899 cm apart and interpenetrated by 1.101 cm.

\begin{figure}[H]
\centering
\includegraphics[width=\linewidth,height=0.75\textheight,keepaspectratio]{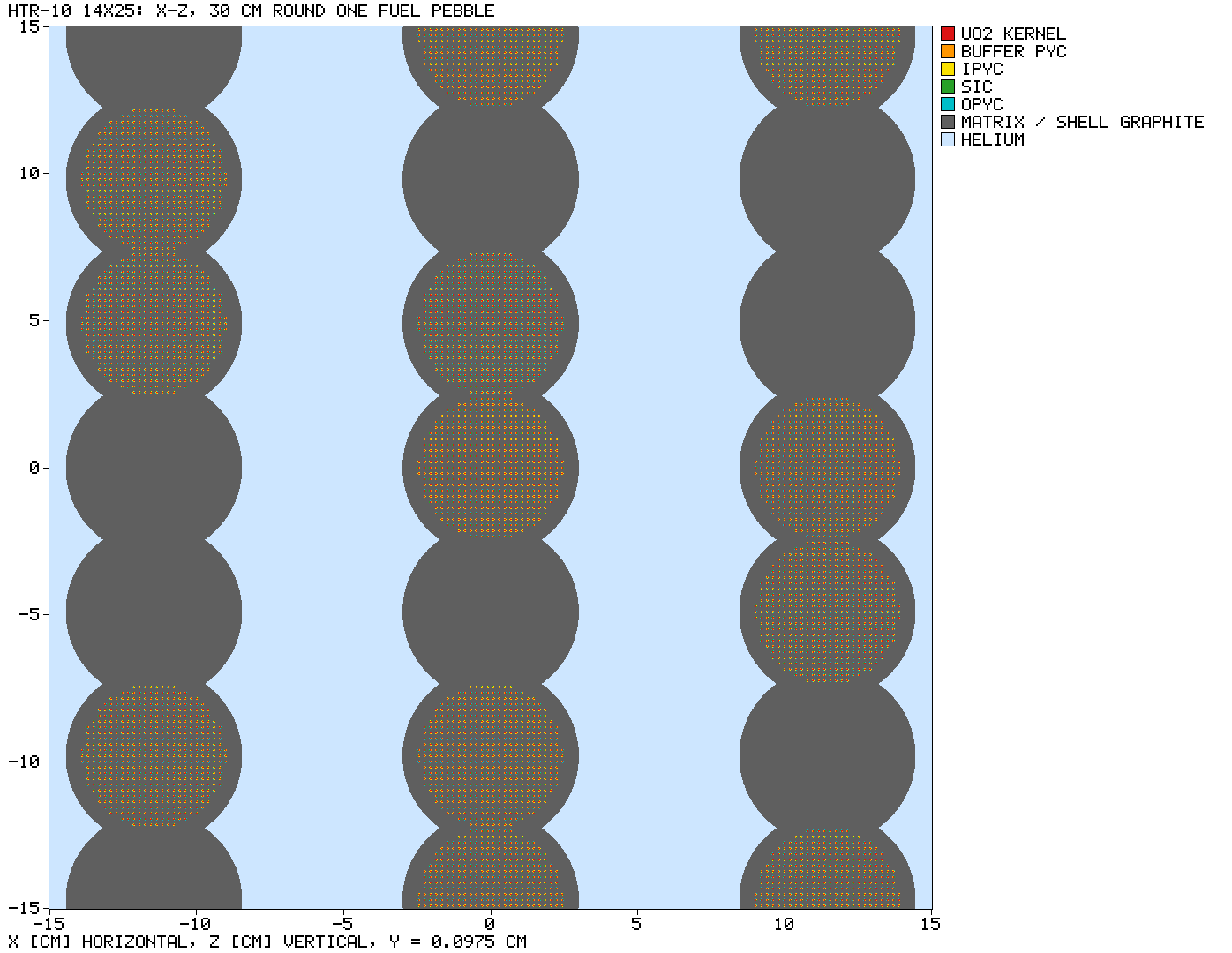}
\caption{Geometry audit of that first HTR-10 pebble-bed model (25 September 2026): x-z slice at y = 0.0975 cm through the fuel pebble at the origin. Pebbles in each column sit 4.899 cm apart and interpenetrate, meeting on a waist with no helium gap. Drawn from the assembled model by the Rust port of OpenMC's slice plotter (code written with AI assistance, Claude Code).}
\label{fig:htr10_xz_pebbles_interpenetrating}
\end{figure}

In this case, the right results were produced with the wrong model, which invalidates the V\&V comparison. The lesson learnt was that it is more important to get the physics right first and the results right second, and that results count only when they rest on physically sound models (which should at least be audited by a human). This supports the view that AI must be driven by human developers with a systematic methodology for producing scientifically correct results, rather than by a brute-force ``just get me this result'' prompt issued as a goal command. The systematic methodology presented in this work was therefore not developed \emph{a priori}, but empirically and \emph{a posteriori}, from lessons learnt from failures encountered during the agentic development and V\&V of Outram Park.

The same principles of differential testing and test-driven development \cite{erdogmus2005testfirst} were used, but their implementation was refined over the course of Outram Park's development. Accordingly, a second, more rigorous method was adopted: agents were instructed to compile the upstream codes, NJOY and OpenMC, and to compare the ports against them, first individually and then together. Specifically, the LLM agents were told not to calibrate simulations to reach answers (unless specifically instructed to), but to put correct processes and physics ahead of correct answers. Only then can results be considered valid. This follows the NRC's principle of unit testing first and integration testing second \cite{nureg_br_0167}. Here, agentic unit testing corresponds to correct processes, and correct integration testing (and qualification testing) corresponds to correct answers.

For unit testing, agents were told to compare the port function by function against the upstream code. The first author did not take part in creating or tracking these unit tests, as there were too many. This allowed the first author to use AI agents to translate code accurately without examining the physics of each function in depth. The agents only needed to replicate each function's behaviour in another language and quantify the difference. If the difference was negligible, i.e. exactly zero or within floating-point error, the translation was considered successful. Machine epsilon is about \(1.2\times10^{-7}\) for f32 and \(2.2\times10^{-16}\) for f64. The latter is typically used for CPU computation, whereas the former is relevant for GPU computation, since GPU shaders (functions running on the GPU, written in languages such as WGSL) often use f32.

Floating-point tolerances may be too stringent in some cases, because rounding differences can arise from numerics and physics that are both correct; such a translation should still count as valid. Judging this normally requires an expert eye. Nevertheless, in the first rounds of agentic translation, debugging and V\&V, coding agents were able to identify defects in the NJOY port's ACE output systematically. The following table is an AI-generated summary of the differences found by a word-by-word survey on 14 October 2026 (Outram Park issue \#875).

\begingroup\small
\setlength{\tabcolsep}{3pt}
\begin{longtable}{>{\raggedright\arraybackslash}p{0.233\linewidth}>{\raggedright\arraybackslash}p{0.233\linewidth}>{\raggedright\arraybackslash}p{0.233\linewidth}>{\raggedright\arraybackslash}p{0.233\linewidth}}
\toprule
\textbf{Nuclide(s)} & \textbf{Difference identified by agentic comparison} & \textbf{Magnitude / extent} & \textbf{Relevance} \\
\midrule
\endhead
Na-23, Cr-50/52/53/54 & Secondary-photon energy data shorter and photon yields differ & 2--6 words shorter; yields differ by \(5\times10^{-6}\) to \(8\times10^{-4}\) & Photon data; not used in neutron transport \\
Fe-57, Cu-65 & Photon-production values differ & 2--4 words differ around the 7th significant figure & Photon data; not used in neutron transport \\
Ti-48 & Total and elastic cross sections differ at a few energies near 6.7 keV & Approximately \(3\times10^{-8}\) relative & Rounding-level difference \\
Zn-64, Zn-67 & Rust port introduces an additional energy point immediately below the unresolved-resonance region & At 0 K, one additional point; after Doppler broadening at 293.6 K, differences extend over approximately 20--28 nearby energies, reaching 0.16\% in total and 0.8\% in absorption & Neutron-data discrepancy requiring investigation \\
Zn-64, Zn-66, Zn-68, Zn-70 & Unresolved-resonance probability tables differ at 0.35 MeV & Capture and total values differ by approximately 1\%; some rarely sampled entries differ by up to 59\%; other table energies identical & Localised unresolved-resonance discrepancy \\
Fe-58 & Unresolved-resonance probability-table range terminates at 1 MeV instead of the 3 MeV upper bound present in the ENDF evaluation and NJOY2016 output & 1--3 MeV region lacks the corresponding probability-table treatment & Significant neutron-processing defect requiring correction \\
Fe-58 & MT=4 total inelastic reaction absent & Reaction omitted from Rust-written ACE file & Potential compatibility issue; effect depends on neutron-transport consumer \\
All examined Rust-written tables & Heating column populated whereas the comparison NJOY2016 output contains zero values & Systematic output difference & Does not affect \(k_\mathrm{eff}\) \\
\bottomrule
\end{longtable}
\endgroup

As the human driver, one can see, even with limited knowledge of the program flow, that the port did not yet produce acceptable ACE files for some nuclides and that further debugging was needed. Here, some expert knowledge is required to prioritise debugging effort. For example, basic nuclear physics coursework should tell the human driver that a cross section misaligned with upstream NJOY is wrong, and that an unresolved-resonance range should not end abruptly where NJOY2016's does not. Upon seeing these defects, agents were assigned to correct them until parity was reached. All rows except the last were fixed on 15 October 2026 (Outram Park commit \texttt{50911476b7}), after which all 76 tables in the survey (38 nuclides at 0 K and 293.6 K) matched NJOY2016 word for word. The heating column differs by design: the NJOY2016 comparison deck runs no HEATR, and heating does not enter \(k_\mathrm{eff}\).

As of October 2026, the Rust port of NJOY had more than 600 agentically generated unit tests (688 in the source tree, and 1148 tests including integration tests). For audit purposes, it was impractical to check every one. However, the key integration test for NJOY is the generation of ACE files, in particular for U-234, U-235 and U-238. The passing criterion was that each ACE file be indistinguishable from the one generated by NJOY2016, down to the last printed digit. This was checked by an agentically generated integration example. In the example, every header, NXS, JXS and XSS entry of the ACE table was compared against the NJOY2016 output as printed, i.e. to 12 significant digits (the precision of NJOY's Type-1 ACE format). Where differences arose, they were printed by the example, and the agents captured this text output to debug the underlying code. The AI agents were instructed to attain parity with NJOY in generating ACE files for U-234, U-235 and U-238, which was achieved between September and October 2026. This check runs as an example rather than as a test, because it would take too long for a regression suite on GitHub Actions and would require the NJOY reference ACE files to be included in the test; for U-235 at 0 K and 293.6 K, these are 316.2 MB and 135.9 MB respectively. Users can, however, instruct agents to rerun the examples at any time.

A similar process was performed for thermal scattering law ACE files, where the outputs of the port were compared against those of upstream NJOY2016. These were made identical, word for word, for Al-27, H in H\(_2\)O, crystalline graphite and H in ZrH. The port's LEAPR output agreed with NJOY2016's to round-off (about \(10^{-13}\) relative). This shows that these parts of the port were translated correctly. ACE files containing cross-section and thermal scattering data can also be checked using the web demos on the GitHub Pages site, which let end users generate ACE files quickly and compare them against those produced by NJOY.

For OutramMC, unit tests were also generated agentically. For integration testing, however, the benchmarks were used, so the integration tests checked for OutramMC also serve as qualification tests. Agentically implemented integration tests were also present, but they were too numerous to check manually.

\subsubsection{Qualification Testing}

Under NUREG/BR-0167, qualification testing is the formal demonstration that software satisfies its baseline requirements \cite{nureg_br_0167}. For the Rust port of NJOY and OutramMC, this means that both the nuclear data processing and the neutron transport capabilities adequately mirror their upstream counterparts. In this work, ``adequately'' means that code-to-code verification is accomplished and that validation is accomplished against selected benchmark cases.

For validation, a benchmark case is successful if it agrees to within the experimental uncertainty. For example, Godiva has a 100 pcm experimental uncertainty band \cite{mcspaden2025music}, and some other criticality benchmarks have experimental uncertainties of around 110--120 pcm. Hence, a 100 pcm uncertainty is tolerable. For the MUSIC criticality benchmarks, calculations were considered to be in good agreement with experiment if they agreed to within 200 pcm \cite{mcspaden2025music}. This provides a useful reference for what is tolerable in the context of validation. However, a 100 pcm tolerance is not universally applicable across all benchmarking and V\&V efforts. Table~\ref{tab:vv_tolerances} lists typical uncertainty and agreement scales for verification and validation:

\begingroup\small
\setlength{\tabcolsep}{3pt}
\begin{longtable}{>{\raggedright\arraybackslash}p{0.315\linewidth}>{\raggedleft\arraybackslash}p{0.315\linewidth}>{\raggedright\arraybackslash}p{0.315\linewidth}}
\toprule
\textbf{Verification / validation context} & \textbf{Typical uncertainty or agreement scale} & \textbf{Reference} \\
\midrule
\endhead
Code-to-code Monte Carlo verification (RMC vs MCNP, C5G7) & Within \(2\sigma\); within 15 pcm & Gao et al. (2013) \cite{gao2013rmc} \\
Godiva (HEU-MET-FAST-001) & $\pm$100 pcm evaluated benchmark uncertainty & McSpaden et al. (2025) \cite{mcspaden2025music} \\
Jemima (IEU-MET-FAST-002) & $\pm$300 pcm evaluated benchmark uncertainty & Mosteller \cite{mosteller2010expanded}; van der Marck \cite{vandermarck2006benchmarking} \\
LCT-008 case 1 & $\pm$160 pcm evaluated benchmark uncertainty & Mosteller \cite{mosteller2010expanded} \\
HST-009 case 1 & $\pm$430 pcm evaluated benchmark uncertainty & van der Marck \cite{vandermarck2006benchmarking} \\
MUSIC critical benchmarks & $<$200 pcm described as good agreement & McSpaden et al. (2025) \cite{mcspaden2025music} \\
Fast-metal ICSBEP benchmarks & Individual uncertainty contributions commonly $<$100 pcm, with larger outliers & Cutler et al. (2023) \cite{cutler2023analysis} \\
KUCA light-water-moderated/reflected HEU cores & 400--742 pcm calculation-to-experiment bias & Pyeon \& Morioka (2022) \cite{pyeon2022uncertainty} \\
KUCA light-water-moderated/reflected HEU cores & \textasciitilde{}919--945 pcm propagated nuclear-data uncertainty & Pyeon \& Morioka (2022) \cite{pyeon2022uncertainty} \\
MCFR full-core calculation & \textasciitilde{}1400 pcm \(k_\mathrm{eff}\) uncertainty from Cl-35 nuclear data, reduced to \textasciitilde{}1000 pcm with the new Cl-35 evaluation & Valentin et al. (2025) \cite{valentin2025uncertainty} \\
\bottomrule
\caption{Typical uncertainty and agreement scales for verification and validation.}\label{tab:vv_tolerances}
\end{longtable}
\endgroup

As Table~\ref{tab:vv_tolerances} shows, Godiva and other fast-spectrum critical-sphere benchmarks typically have uncertainties from below 100 pcm up to at most 200 pcm. For light-water-moderated cores, agreement to within 400--950 pcm was considered acceptable. For full-core calculations of designs such as the MCFR, Cl-35 nuclear data alone contribute about 1000--1400 pcm of \(k_\mathrm{eff}\) uncertainty. For code-to-code verification, however, a tighter \(2\sigma\) criterion is used; Gao et al., for example, found RMC and MCNP eigenvalues within two standard deviations and within 15 pcm of each other \cite{gao2013rmc}. These values define the tolerances within which criticality calculations can be considered to pass.

Neutronics calculations consist of two parts: nuclear data processing, which can be provided by NJOY or other software, and neutron transport, which can be provided by OpenMC or other software. For the Rust port of NJOY to be acceptable, it should reproduce NJOY's cross sections, ideally byte for byte or to within floating-point error, or at least to within some tolerance. However, such a tolerance may be considered too tight in some instances. For the purposes of neutron transport, it may be sufficient that two nuclear data files, one produced by NJOY and one by the Rust port, give statistically indistinguishable results in the Monte Carlo code. Tada et al. used this approach for code-to-code verification of FRENDY against NJOY \cite{tada2017frendy}. When individual processed files are compared directly, a 0.1\% tolerance has been used in the literature, and a 0.01\% tolerance for Doppler broadening and resonance reconstruction \cite{tada2017frendy}. For reference, the typical reconstruction tolerance used in RECONR is about 0.1\% \cite{njoy2016}.

In light of this, differential testing becomes important: ACE files from both NJOY and its Rust port are fed into OpenMC, and the resulting k-eigenvalues are compared. This verifies the ACE file generation and isolates errors in the transport port, OutramMC, from errors in nuclear data generation. Next, OutramMC is made to read ACE files processed by NJOY, which isolates port errors in transport. Once both steps are done, OutramMC can be coupled to the Rust NJOY port in two ways. The first is direct coupling, in which the Rust NJOY port is a dependency of OutramMC and nuclear data are passed through memory without writing a single ACE file. This is useful where reading and writing ACE files is disadvantageous, for example when disk space is short or file input/output is unavailable or cumbersome. The second is to generate ACE files with the Rust port of NJOY and have OutramMC read these files to perform neutron transport.

For qualification testing, agents need a criterion for when to stop refining a solution. An example arose early in diagnosing discrepancies between OutramMC routes 3, 4 and 5 and OpenMC route 1 (the routes are defined below). When trying to determine whether a discrepancy was within statistical uncertainty, the agents kept adding more and more k-eigenvalue runs with different seeds. Here, a seed refers to one full Monte Carlo k-eigenvalue calculation with 5000 or 10000 neutrons per batch, a set of inactive batches discarded while the fission source converges, and a set of active batches over which k-eigenvalue statistics are tallied.

In these calculations, 5000 particles per batch were used for Godiva, Jemima and HST-009, and 10000 particles per batch for LCT-008. This is consistent with Brown's guidance that ``at least 5000 or more neutrons per cycle should be used for long production runs'' to prevent bias in \(k_\mathrm{eff}\) and reaction-rate tallies \cite{brown2009mcreview}. For fission source convergence, there must be enough inactive batches for the fission source to stop changing from batch to batch. This is measured by the Shannon entropy of the fission source, which should reach a roughly constant value (a plateau) \cite{brown2009mcreview}. Once the Shannon entropy flattens out, statistics can be collected. In the Results section, the Shannon entropy is therefore used to quantify fission source convergence and to justify the number of inactive batches.

The AI agents also determined that runs over a sufficient number of seeds were needed to resolve whether discrepancies were due to the different random number streams or to an underlying physics problem. In brute-force agentic debugging, the agents found that OutramMC was about 50--73 pcm below OpenMC at 32 seeds for LCT-008, corresponding to 3.8--4.9 standard deviations, which clearly established a discrepancy.

Yet, in an effort to check thoroughly, the AI agents increased the number of seeds to 96, with 250 inactive and 400 active batches. This is consistent with Brown's guidance that ``it is preferable to use 10,000, 20,000, 50,000, or more neutrons per cycle, as long as a few hundred active cycles are computed'' in order to prevent bias in \(k_\mathrm{eff}\) and reaction-rate tallies \cite{brown2009mcreview}. The larger run resolved the remaining discrepancy to about 18 pcm, or 2.3 standard deviations with a standard deviation of 8 pcm, but it did not show where the discrepancy and the underlying bug came from. These were meant to audit the effect of fixes made to the angular sampling of the correlated energy--angle laws (ENDF laws 44 and 61), now performed as in OpenMC; to the unresolved resonance range (URR) probability tables, now sampled with one band per nuclide and energy; to the energy gate and window that decide where the Doppler broadening rejection correction (DBRC) is applied; and to the sampling of the thermal scattering law \(S(\alpha,\beta)\), now equiprobable sampling as in OpenMC. However, this was an ineffective solution. To ensure parity with upstream, it was more computationally efficient, both in Monte Carlo runs and in tokens, to ensure ACE file parity with NJOY2016 before Monte Carlo runs were performed.

For the simple goal of finding out whether the two codes disagreed, this may have been excessive, as the larger run only confirmed what the lower-statistics run had already established. In hindsight, it would have been more efficient to accept the low-statistics run as confirmation of a discrepancy and to search for its cause systematically. Moreover, after the discrepancy was resolved, the LCT-008 run became part of regression testing. Left at these high statistics, it took almost half a day on a desktop when left unchecked, whereas the other three cases each finished in under an hour. For regression testing, we only require that the code still functions sufficiently well after changes to the codebase. Here, ``sufficiently well'' means that the simulation runs as a whole and produces k-eigenvalues statistically indistinguishable from OpenMC's. For such runs, 96 seeds of 250 inactive and 400 active batches at 10000 particles per batch may have been excessive.

For comparison, in code-to-code verification of RMC against MCNP, Gao et al. ran 50,000 particles per generation over 1500 cycles, with the number of inactive cycles chosen with the help of the Shannon entropy, and found the RMC and MCNP eigenvalues to differ by 1--15 pcm (their Table 2) \cite{gao2013rmc}. This amounts to at most 75 million active neutron histories per case. In contrast, 32 independent seeds of 40 inactive and 120 active batches at 5000 histories per batch (10000 for LCT-008) would have sufficed for code-to-code verification here. This amounts to about 19.2 million active histories (38.4 million for LCT-008).

The agents chose independent seeds to minimise correlation between results: each generation depends on the fission bank of the previous one, so batches within one run are correlated, and Monte Carlo codes that ignore this inter-cycle correlation underestimate the true uncertainty. Splitting the calculation across seeds makes the runs truly independent, rather than averaging correlated batches within a single run. Brown recommends this approach as a brute-force way of obtaining the true uncertainties: ``make 25 or so independent Monte Carlo criticality calculations, discarding the uncertainties from the individual calculations, and compute the true uncertainties from the ensemble of results'' \cite{brown2009mcreview}. The approach is slightly more expensive per history, because the fission source must converge, as judged by the Shannon entropy, before tallying starts in each of the 32 independent runs. Overall, it uses fewer neutron histories per calculation and therefore gives larger standard deviations. For ascertaining whether the port has discrepancies, however, it is a good compromise between computation time on a desktop and accuracy, and it is therefore suitable for an iterative approach. The approach was initially adopted by the agents and later refined by the human developer.

The more effective approach was to compare against upstream OpenMC and NJOY with the number of seeds fixed at 32 and the k-eigenvalue settings fixed in advance. This did not require in-depth knowledge of how to write OpenMC or NJOY from scratch, but it did require a basic understanding of where bugs typically arise in such software. The developer using AI therefore needs some prior experience in running neutronics Monte Carlo codes, to prevent agents from running unnecessarily long calculations. In qualification testing, it was desired to isolate the effects and porting bugs of nuclear data software and that of the transport code. Therefore, the qualification-testing methodology consists of five routes:

\begin{enumerate}
  \item OpenMC + NJOY2016 ACE (converted to HDF5)
  \item OpenMC + Rust-NJOY ACE (converted to HDF5)
  \item OutramMC + NJOY2016 ACE
  \item OutramMC + Rust-NJOY ENDF (direct coupling, no ACE file written)
  \item OutramMC + Rust-NJOY ACE
\end{enumerate}

\begin{table}[H]
\centering
\caption{Five-route qualification-testing matrix: nuclear data processing (rows) against Monte Carlo neutron transport (columns).}
\label{tab:five_route_matrix}
\begin{tabular}{lcc}
\toprule
 & OpenMC & OutramMC \\
\midrule
NJOY2016 & Route 1 & Route 3 \\
Outram Park's NJOY fork & Route 2 & Routes 4 and 5 \\
\bottomrule
\end{tabular}
\end{table}

For these five routes, validation is considered successful if the \(k_\mathrm{eff}\) values fall within the applicable evaluated benchmark uncertainty: 100 pcm for Godiva, 300 pcm for Jemima, 160 pcm for LCT-008 case 1 and 430 pcm for HST-009 case 1. Code-to-code verification is successful if the routes agree within \(2\sigma\) of each other, and this should be within 50 pcm. This was decided as a suitable compromise between computational cost and accuracy for several agentic test runs to make successive iterations faster. This is because regression tests were often rerun by agents, at the first author's instruction, after changes in code to test if the agentic changes in code caused performance degradation in the nuclear data and Monte Carlo runs. For most code changes, it was not practical to have runs exceeding 12 hours just to test the effect for a change in one line of code.

\subsubsection{Acceptance Testing}

Acceptance testing establishes that the software works in its intended environment. Here, the pure Rust ports of NJOY and OpenMC (OutramMC) should run natively from the command line on Windows, macOS and Linux, including Termux on Android phones. iOS is most conveniently reached through web demos, in which Rust is compiled to WebAssembly and coupled with a graphical user interface library, giving the user an interactive, user-friendly way to try the cases in a web browser. There, the nuclear data must be processed in the browser, and the validation results must also be produced in the browser, on the same cases used for validation.

This can be done through continuous integration \cite{slaughter2021continuous,soares2022continuous} and by publishing to GitHub Pages via GitHub Actions. Acceptance testing for the NJOY port and OutramMC is therefore considered complete only once both the CI and the web demos pass. Running high-fidelity cases in CI to within 50 pcm would take too long to be practical, so CI uses a 200 pcm allowance against route 1 (OpenMC + NJOY2016), at 5000 particles per batch with 40 inactive and 120 active batches. In the web demos, users can choose the number of particles, inactive batches and active batches through a graphical user interface, just as they would by editing an OpenMC settings file. These criteria were met on 13 October 2026 (Outram Park issue \#859): CI runs all four cases on Linux, Windows, macOS and an Android emulator, and the web demos run all four cases in the browser, including on a phone.

Web demos have the added advantage of making results easy to reproduce. The user only has to load the webpage: the benchmark case comes with its settings preloaded, along with the random seed used for the test, and the ENDF files are processed in the browser by the Rust NJOY port, which takes a few minutes. Thereafter, with a click of the mouse, the user can start running the benchmark case. This makes reproducing the benchmark cases accessible to anyone, even those who do not know how to compile the code.

Web demos can also satisfy the traceability requirement of NUREG/BR-0167 \cite{nureg_br_0167}. Traceability requires that the exact version of the code used for a benchmark be tracked, which is given by the git commit and the date on which the \(k_\mathrm{eff}\) run was generated. The web demos for these benchmarks were therefore designed to display the commit and date of each recorded benchmark run alongside the live results of the Monte Carlo runs initiated by the user. This lets the user see clearly at which commit and on which date the recorded runs were performed.

\subsection{Documentation and Deliverables}

Under NUREG/BR-0167, documentation ensures that the resulting software can be understood, operated, maintained and audited \cite{nureg_br_0167}. For these purposes, the user first needs an idea of what the software is about and which scientific concepts it implements. This matters even for human-written code. In agentic software development, however, software engineering practices are moving away from human-readable code \cite{ustynov2026beyond,monperrus2026codereview}. Yet even in the age of agentic software development, as Minister for Foreign Affairs Dr Vivian Balakrishnan emphasised in a speech to AI Engineer Singapore on 16 May 2026, understanding and accountability remain fundamentally human responsibilities. He remarked that although calculations, computation, memory and other tasks can increasingly be outsourced, personal understanding cannot be outsourced, and accountability cannot be delegated \cite{balakrishnan2026aiengineer}. Therefore, whichever agentic tools are used, understanding and accountability cannot be delegated to agents. Even as software engineering practices evolve in 2026 with the advent of AI agents, it remains the responsibility of the human directing the agents to understand the software and to remain accountable for its correctness.

This can be achieved through thorough documentation and selective reading of the portions of source code that are important to understanding how the software functions. Huang and Madeira proposed targeted code inspection in 2022, in which inspection effort is directed toward portions of software considered more important or error-prone \cite{huang2022targeted}. Similarly, Baltes et al. (2017) observed how developers navigate and inspect relevant portions of software when investigating performance problems \cite{baltes2017navigate}. These practices demonstrate that inspecting and improving a codebase does not necessarily require every line of source code to be read and understood. Importantly, these approaches predate the agentic AI coding era. Selective code reading, of course, applies only to the portions of the code relevant to the task under investigation. For agentic nuclear research software development, it is paramount that (1) the physics is accurately implemented in code, within reasonable computational costs and numerical bounds, and (2) the code remains sufficiently readable and understandable to be audited and used even without AI agents. These principles are consistent with the expectations of software quality assurance and should continue to be followed.

In agentic coding, reading the resulting code can resemble exploring a new open-source codebase. The human directing agentic software development encounters the repository much as a new user encounters the software for the first time. Consequently, the developer's own inspection provides an opportunity to audit whether the codebase can be understood by other users. Human review of the source code and its documentation therefore serves two purposes: (1) it assesses whether the codebase is sufficiently readable and understandable for users, and (2) it assesses whether the codebase implements the intended scientific concepts correctly. The first point of contact with such a codebase is often its documentation, including for agentically generated software. Indeed, Yuan et al. (2026) propose using ``unambiguous Natural Language (NL)-based documentation as a readable intermediate layer between prompts and code'' \cite{yuan2026verifiable}.

Within the documentation, the first thing a human reader (or the human driving the AI agents) wants to know about nuclear science software is what physics it implements. For this, tutorial-style documentation with web demos is helpful, because humans may find it easier to interact with a web demo than to read many lines of documentation. Such practices were explored and implemented in Casdoc, where users interacted with functional portions of code in non-linear documentation rather than reading code line by line \cite{casdoc2025}. This supports the use of an interactive environment in which users interact with code, in a course-like manner where readers, like students, work through coursework and programming examples in the browser. However, Casdoc assumes that readers know how to code (or want to). A reader new to the codebase may first want to understand the physics without having to code, before moving on to the software implementation of that physics.

Therefore, web demos need not be coding studios in the browser; they can be interactive demonstrations of the physics the code implements. In physics education, Wieman et al. showed that interactive simulations support engagement and conceptual learning \cite{wieman2008interactive}. Web demos can therefore be used to explain the physics concepts used in the code. In Outram Park, web demos serve two purposes: (1) to prove that the code runs in the browser, as part of acceptance testing, and (2) to provide interactive demonstrations that help the user understand the code, as part of documentation. Since web demos can be written quickly with AI agents, several of them can be used to produce effective code documentation. This approach is used for Outram Park, and for OutramMC in particular.

\section{Results}

\subsection{Agentic Translation and Differential Testing}

\subsubsection{Unit and Integration Testing of the NJOY Rust Port}

Based on the methodology, agentic V\&V was performed using the Rust fork of NJOY and OutramMC as an example. In initial tests, U-238 Doppler broadening was used as an initial visual test of whether the Doppler broadening ports produced correct behaviour.

\begin{figure}[H]
\centering
\includegraphics[width=\linewidth]{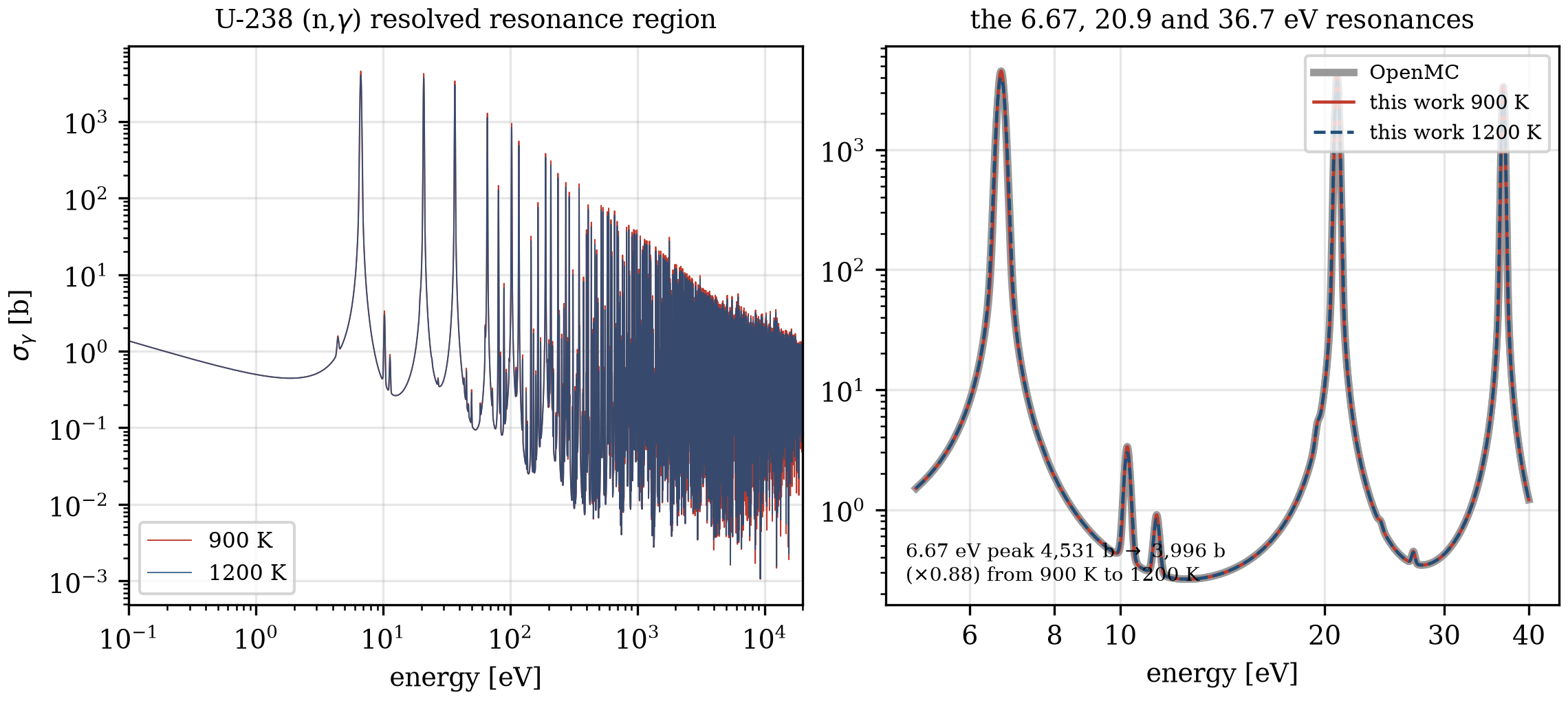}
\caption{U-238 radiative capture (MT=102) across the resolved resonance region, reconstructed by the Rust port of NJOY from the ENDF/B-VIII.0 evaluation and Doppler-broadened to 900~K and 1200~K. Data visualisation produced by a plotting script (\texttt{make\_figures.py}) written with AI assistance (Claude Code).}
\label{fig:u238-forest}
\end{figure}

\begin{figure}[H]
\centering
\includegraphics[width=\linewidth]{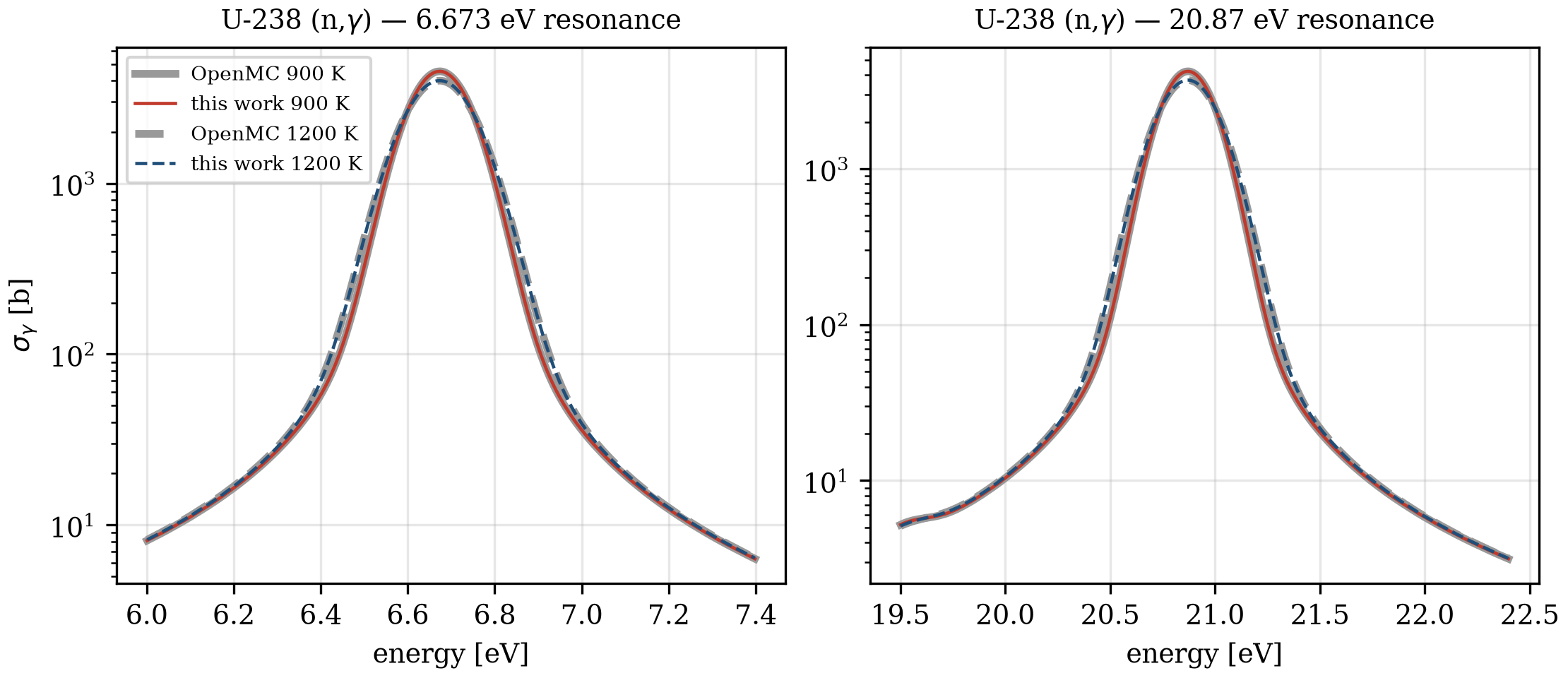}
\caption{The 6.673~eV and 20.87~eV resonances at both temperatures, with this work drawn over the OpenMC reference. Data visualisation produced by a plotting script (\texttt{make\_figures.py}) written with AI assistance (Claude Code).}
\label{fig:u238-resonances}
\end{figure}

\begin{figure}[H]
\centering
\includegraphics[width=\linewidth]{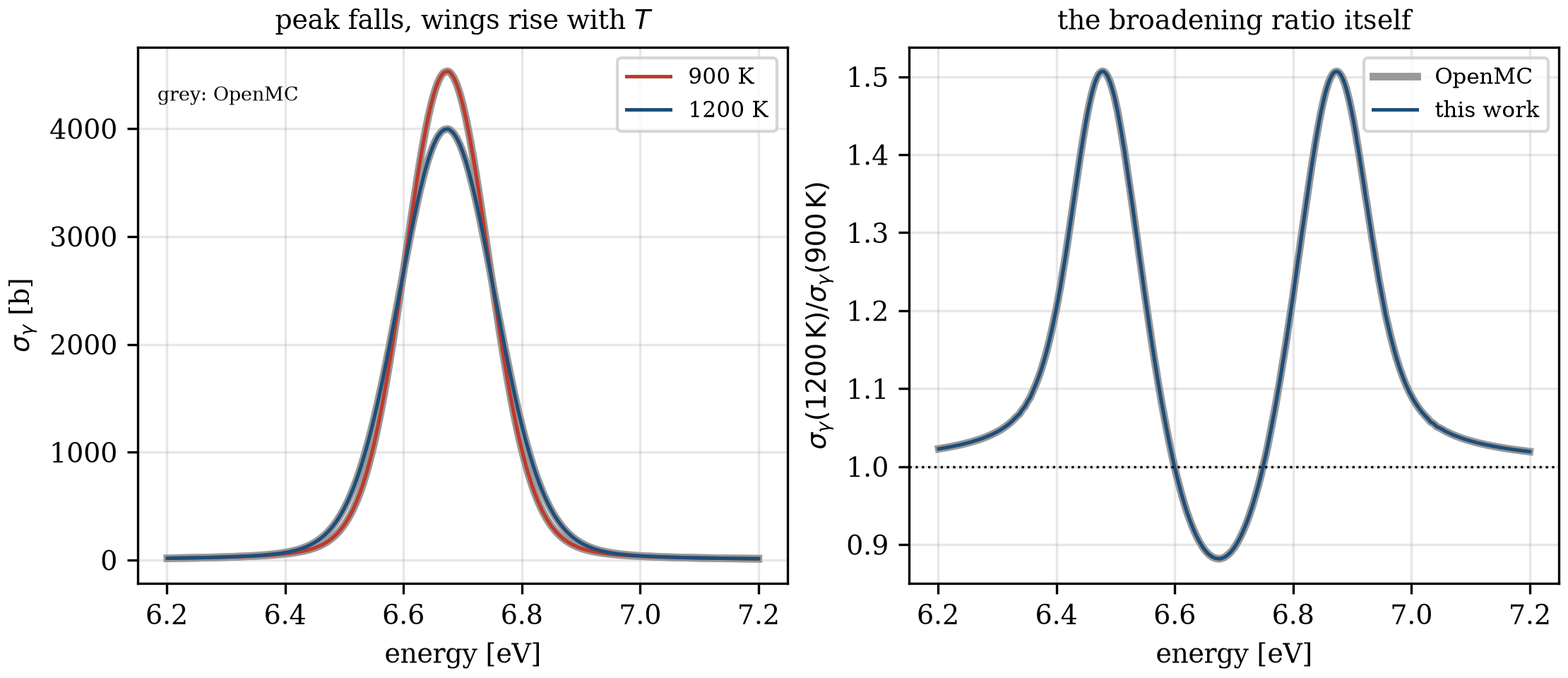}
\caption{\emph{Left:} the broadening itself: the 6.673~eV peak falls and its wings rise with temperature, with OpenMC in grey. \emph{Right:} the ratio $\sigma_\gamma(1200\,\mathrm{K})/\sigma_\gamma(900\,\mathrm{K})$, computed independently within each code. Data visualisation produced by a plotting script (\texttt{make\_figures.py}) written with AI assistance (Claude Code).}
\label{fig:u238-broadening}
\end{figure}

U-238 is of particular interest because it is primarily responsible for the negative fuel temperature feedback vital to reactor safety. U-238 radiative capture cross sections at two temperatures, 900~K and 1200~K, were therefore reconstructed and Doppler-broadened by the Rust port of NJOY and compared point for point against OpenMC's pre-generated HDF5 library, which was itself produced by NJOY from the same ENDF/B-VIII.0 evaluation. The modules exercised are RECONR and BROADR, which reconstruct the resonances and Doppler-broaden them. This is therefore a code-to-code verification case, shown in Figures~\ref{fig:u238-forest}, \ref{fig:u238-resonances} and \ref{fig:u238-broadening}.

After these encouraging initial verification results, it was then decided that the codebase could then proceed onto systematic unit testing. In the course of unit testing, more than 600 unit tests were generated for NJOY and OutramMC when agents were told to construct unit tests function by function. These were not manually inspected. Agents reported parity for U-234, U-235 and U-238 ACE files generated by the NJOY Rust port, where every value was the same down to the last decimal point. This is shown in the following agentically reported table:

\begingroup\small
\setlength{\tabcolsep}{3pt}
\begin{longtable}{>{\raggedright\arraybackslash}p{0.184\linewidth}>{\raggedleft\arraybackslash}p{0.184\linewidth}>{\raggedleft\arraybackslash}p{0.184\linewidth}>{\raggedleft\arraybackslash}p{0.184\linewidth}>{\raggedleft\arraybackslash}p{0.184\linewidth}}
\toprule
 & \textbf{U-234, 293.6 K} & \textbf{U-235, 293.6 K} & \textbf{U-235, 0 K} & \textbf{U-238, 293.6 K} \\
\midrule
\endhead
Processing chain & RECONR $\rightarrow$ BROADR $\rightarrow$ PURR $\rightarrow$ ACER & RECONR $\rightarrow$ BROADR $\rightarrow$ PURR $\rightarrow$ ACER & RECONR $\rightarrow$ ACER & RECONR $\rightarrow$ BROADR $\rightarrow$ PURR $\rightarrow$ ACER \\
Energy grid points & 25,393 & 76,027 & 233,515 & 155,207 \\
Grid points in common with NJOY & 100\% & 100\% & 100\% & 100\% \\
Reactions & 49 & 85 & 84 & 49 \\
Fission $\bar{\nu}$ (NU block) & 37 words & 347 words & 347 words & 181 words \\
Probability tables (LUNR) & 3,152 words, 26 energies & 3,152 words, 19 energies & Absent in both & 3,152 words, 83 energies \\
Whole table & 449,695 words & 6,712,632 words & 15,616,079 words & 6,247,445 words \\
Header arrays (NXS, JXS) & Same & Same & Same & Same \\
Rust build time & 16 s & 63 s & 16 s & 49 s \\
\bottomrule
\end{longtable}
\endgroup

This table provides agentically generated evidence that the NJOY port for ACE file generation was successful for the nuclides mentioned. Of course, ACE file parity was still required for LEAPR ports.

For a start, LEAPR files from the NJOY port were compared against the thermal scattering law files in ENDF. This ensured that the LEAPR module itself showed parity. Agentically generated reports show that the port was accurate:

\begingroup\footnotesize
\setlength{\tabcolsep}{3pt}
\begin{longtable}{>{\raggedright\arraybackslash}p{0.111\linewidth}>{\raggedright\arraybackslash}p{0.111\linewidth}>{\raggedright\arraybackslash}p{0.111\linewidth}>{\raggedright\arraybackslash}p{0.111\linewidth}>{\raggedright\arraybackslash}p{0.111\linewidth}>{\raggedright\arraybackslash}p{0.111\linewidth}>{\raggedright\arraybackslash}p{0.111\linewidth}>{\raggedright\arraybackslash}p{0.111\linewidth}}
\toprule
\textbf{Case} & \textbf{MAT} & \textbf{T (K)} & \textbf{Quantity compared} & \textbf{Points} & \textbf{Worst rel. dev. (recorded)} & \textbf{Re-run 2026-10-14} & \textbf{Gate} \\
\midrule
\endhead
Graphite, MT4 kernel & 30 & 296 & unrounded kernel S vs stored S, $>$ 1e-30 & 48,941 of 60,000 (6,645 stored zeros exact) & 4.917e-6 max, 6.390e-7 RMS (pre-2017 k\_B); 3.711e-4 max (CODATA2018 k\_B) & same & max 1e-5, RMS 2e-6, zero pattern identical \\
Graphite, MT4 kernel & 30 & 1000 & as above & 52,378 (6,328 zeros exact) & 4.838e-6 max, 5.817e-7 RMS & same & max 1e-5, RMS 2e-6, zero pattern identical \\
Graphite, MT4 end-to-end & 30 & 296 & stored S, $>$ 1e-30 & 48,941 (6,645 bit-identical) & 1.004e-13 max (2026-09-11); 1.002e-13 max, 9.993e-14 RMS (re-measured 2026-10-04) & not re-run & max $<$ 1e-12 \\
Graphite, MT2 coherent elastic & 30 & 10 T, 296--2000 & Bragg edges + cumulative S(E;T) & 221 edges; 2,200 S(E;T) & edges 9.937e-7; S(E;T) 9.986e-7 max, 2.408e-7 RMS (CODATA2018); 1.001e-13 with the pre-2017 constants & not re-run & same edge count; max 2e-6, RMS 1e-6 \\
Si in SiC, MT4 kernel & 43 & 8 T, 300--1200 & MF7/MT4 S $>$ 1e-12 & 306,033 & 0.0000 \% (to the 6 figures ENDF stores) & not re-run & 1 \% per point \\
SiC coherent elastic, 0.0253 eV & 43; 44 & 296 & $\sigma$\_coh,el & 1 per material & \textbf{$-$2.96 \% to $-$3.03 \%} (a characterised disagreement) & not re-run & within 5 \% of the tape \\
\bottomrule
\end{longtable}
\endgroup

Of course, for the LEAPR generated files in this table, one must decide whether the errors are acceptable. For example, in graphite, for the MT4 kernel, is an error of 4.838e-6 acceptable? This requires human judgment. Whereas for the U-234, U-235 and U-238 files, exact reproduction where every number matches that of NJOY eliminates the need for human judgment. However, the trade-off is that more time, tokens and computing power are required to produce ACE files to that precision. Hence, it is evident that the trade-off here is that a higher-fidelity port requires less human judgment. Where the port provides exactly the same files as upstream when comparing every value in the ACE table (or output datafile), then it is easy for humans, even without expertise, to judge if the port is successful. This option is possible provided there are upstream datafiles one can use as an oracle to replicate so as to judge the success of the port.

For agentically implemented code without such an upstream port, expert human judgment is required to check if error tolerances are acceptable. For example, in the last row, SiC coherent elastic had significant differences from the ENDF files, about 3\%. This was because open-source NJOY2016 did not have a way to process LEAPR input for SiC. The LEAPR code was an extension agentically implemented from Zhu, Y. (2014), \emph{Thermal Neutron Scattering Cross Sections for Silicon Carbide}, MS thesis, NC State, supervised by Hawari (NCSU handle 1840.16/9396) \cite{zhu2014thermal} by sending the paper into Claude and asking it to implement a draft for SiC. There was no upstream to compare against. However, a correct implementation should replicate the ENDF file. As to whether a 3\% error is acceptable, that requires human judgment. This SiC LEAPR implementation is currently work in progress, but outside the scope of this work. SiC is important because it forms one of the most important protective layers in TRISO particles.\footnote{For neutron scattering, it is important to have the correct thermal scattering laws for SiC. However, in older simulations of HTR-10 using ENDF/B-VII.0 and before, there were no thermal scattering laws for SiC in ENDF libraries. These came only with later libraries such as ENDF/B-VIII.0. Li et al. \cite{li2014research} used free-gas SiC scattering in the HTR-10 model then, and that was the best they could do given the data. One could argue that the impact on neutron moderation is small as most moderation is done within graphite rather than SiC. However, the first author thought it was worthwhile to attempt reimplementing the LEAPR processes for SiC in the Rust port as it only cost some tokens and a few hours of agentic coding and implementation.} This remains future work where knowledgeable collaborators can be involved. Moving on, the agentically generated evidence for thermal scattering law ACE file parity is shown in the following table:

\begin{table}[H]
\centering
\footnotesize
\setlength{\tabcolsep}{3pt}
\caption{Thermal scattering law ACE files produced by the Rust port of NJOY, compared with those of NJOY2016.}
\label{tab:tsl_ace_parity}
\begin{tabular}{>{\raggedright\arraybackslash}p{0.086\linewidth}>{\raggedright\arraybackslash}p{0.086\linewidth}>{\raggedright\arraybackslash}p{0.086\linewidth}>{\raggedright\arraybackslash}p{0.086\linewidth}>{\raggedright\arraybackslash}p{0.086\linewidth}>{\raggedright\arraybackslash}p{0.086\linewidth}>{\raggedright\arraybackslash}p{0.086\linewidth}>{\raggedright\arraybackslash}p{0.086\linewidth}>{\raggedright\arraybackslash}p{0.086\linewidth}>{\raggedright\arraybackslash}p{0.086\linewidth}}
\toprule
\textbf{Thermal material (MAT)} & \textbf{Incident (MAT)} & \textbf{T (K)} & \textbf{Tabulated T used (K)} & \textbf{tol} & \textbf{natom} & \textbf{Bins} & \textbf{iwt (IFENG)} & \textbf{Elastic} & \textbf{Result} \\
\midrule
Al-27 (53) & Al-27 (1325) & 20 & 20 & 0.05 & 1 & 16 & 1 (0) & coherent & identical \\
Al-27 (53) & Al-27 (1325) & 20 & 20 & 0.001 & 1 & 16 & 1 (0) & coherent & identical \\
Al-27 (53) & Al-27 (1325) & 20 & 20 & 0.001 & 1 & 16 & 0 (1) & coherent & identical \\
Al-27 (53) & Al-27 (1325) & 20 & 20 & 0.001 & 1 & 16 & 2 (2) & coherent & identical \\
H in H$_2$O (1) & H-1 (125) & 283.6 & 283.6 & 0.001 & 2 & 16 & 1 (0) & none & identical \\
H in H$_2$O (1) & H-1 (125) & 283.6 & 283.6 & 0.001 & 2 & 64 & 0 (1) & none & identical \\
H in H$_2$O (1) & H-1 (125) & 283.6 & 283.6 & 0.001 & 2 & 64 & 2 (2) & none & identical \\
H in H$_2$O (1) & H-1 (125) & 293.6 & 293.6 & 0.001 & 2 & 16 & 1 (0) & none & identical \\
H in H$_2$O (1) & H-1 (125) & 293.9 & 293.6 & 0.001 & 2 & 16 & 1 (0) & none & identical \\
H in H$_2$O (1) & H-1 (125) & 330 & --- & 0.001 & 2 & 16 & 1 (0) & none & refused by both \\
Crystalline graphite (30) & C-12 (625) & 296 & 296 & 0.001 & 1 & 16 & 1 (0) & coherent & identical \\
Crystalline graphite (30) & C-12 (625) & 600 & 600 & 0.001 & 1 & 16 & 1 (0) & coherent & identical \\
Crystalline graphite (30) & C-12 (625) & 600.5 & 600 & 0.001 & 1 & 16 & 1 (0) & coherent & identical \\
Crystalline graphite (30) & C-12 (625) & 450 & --- & 0.001 & 1 & 16 & 1 (0) & coherent & refused by both \\
Crystalline graphite (30) & C-12 (625) & 603 & --- & 0.001 & 1 & 16 & 1 (0) & coherent & refused by both \\
H in ZrH (7) & H-1 (125) & 296 & 296 & 0.001 & 1 & 16 & 1 (0) & incoherent (IDPNC = 3) & identical \\
\bottomrule
\end{tabular}
\end{table}

This shows that the ACE files generated by the Rust port of NJOY were essentially identical to those of NJOY2016.

The parity of ACE files allows non-expert human eyes to judge that the port was relatively successful as the ACE files generated through the Rust port were essentially identical to those produced from NJOY. However, for the LEAPR generated files, there were still some residuals which required human judgment. In such cases, the best way to check from a Monte Carlo practitioner's perspective is to run the five-route calculation, which is the qualification step. This can be done without expert knowledge of LEAPR's algorithms, thus supporting the point that agentic porting and testing can be done successfully and assessed by those without high levels of expertise. Also, once the Rust fork of NJOY reaches parity, it has no further reason to evolve as it essentially reproduces performance of upstream code nearly exactly. At this point, the code converges in design, and limited human reviews need not be wasted on evolving versions of code.

\subsection{Five-Route Qualification Testing}

Given the parity of the ACE files generated by the Rust port of NJOY, it was important to move on to the five-route test mentioned earlier for qualification testing. For five-route testing, Godiva, Jemima, HST009 and LCT008 were used as they were common ICSBEP benchmarks to use for neutronics code validation. The statistics used were 40 rejected batches, 120 batches used for statistics, 5000 neutrons per batch for Godiva, Jemima and HST009, and 10000 neutrons per batch for LCT008. This was taken from guidance of Brown et al, where a minimum of 5000 batches were required to minimize biasing, 10000 batches were used for LCT008 well above the 5000 minimum. These were chosen as compromises between computation speed and accuracy, as several runs of the five-route test were run during the course of agentic debugging. Thirty-two independent seed draws were run for each of these cases. This allowed seeds to be more independent of each other as compared to individual batches within the same random seed. The statistics were collated and presented in Figure~\ref{fig:five_route_keff}:

\begin{figure}[H]
\centering
\includegraphics[width=\linewidth,height=0.75\textheight,keepaspectratio]{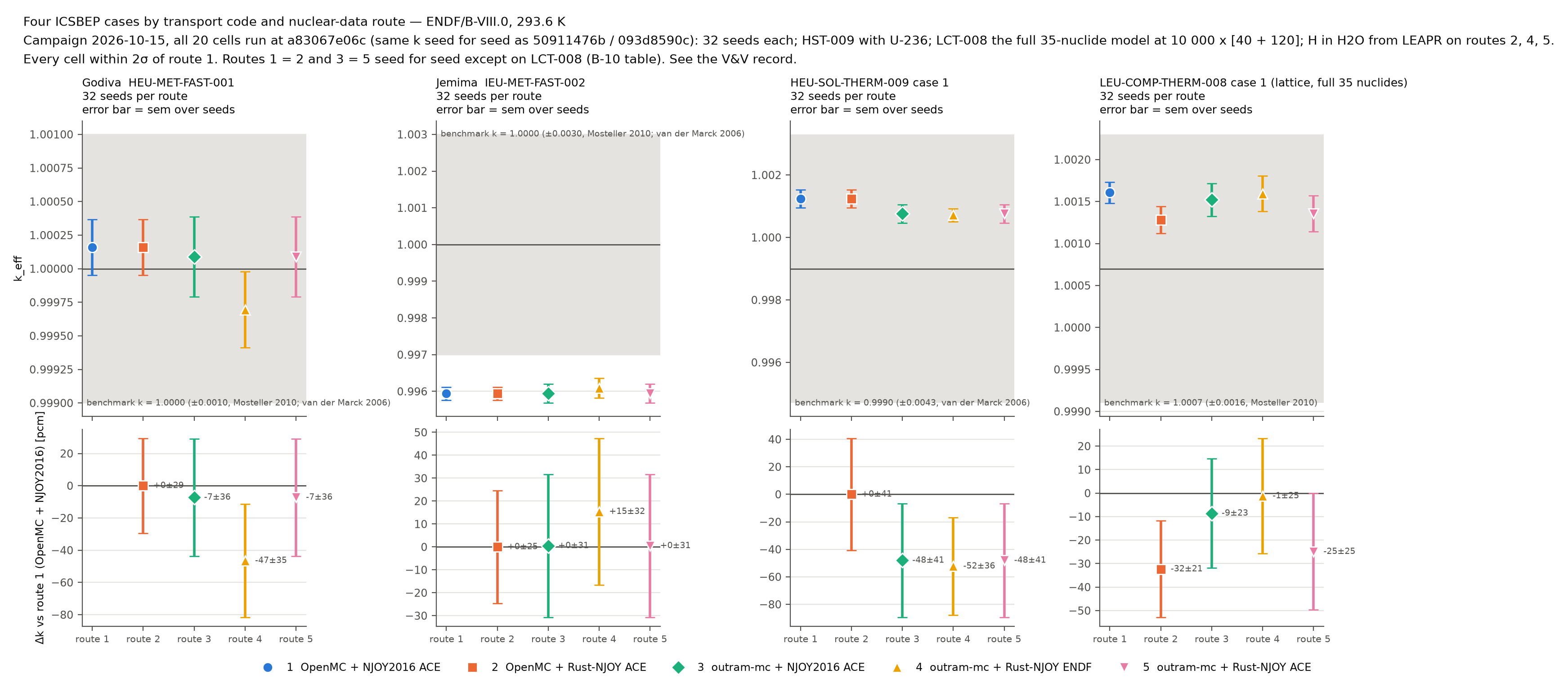}
\caption{Five-route qualification results for Godiva, Jemima, HST009 and LCT008.}
\label{fig:five_route_keff}
\end{figure}

Routes 2--5 show good agreement with route 1 for all cases, agreeing to within 2 standard deviations. For Godiva, Jemima and HST009, routes 1 and 2 were basically identical. This shows that the ACE files generated by the NJOY Rust port were indistinguishable from NJOY2016. Further evidence of ACE file parity can be shown between routes 3 and 5, where NJOY2016 ACE files and Rust's NJOY fork were used in OutramMC.

However, for LCT008, routes 1--2 and routes 3 and 5 were within 2 standard deviations of each other. This was statistically indistinguishable but still a noticeable difference. The reason was rounding errors in B-10. The reason for this, as agents reported, is that NJOY2016's writer crashed on B-10 as it output the ACE files. Therefore, the NJOY2016 B-10 library containing unrounded numbers was used. These were different from the text ACE files for B-10 generated using the Rust fork of NJOY. Agents reported that ``in the HDF5 files OpenMC reads, B-10 differs by at most 1e-13 in every array, and its temperature by 1.9e-5 (text prints it to 4 figures).'' The temperature difference was as follows according to agentic summaries. ACE files store temperatures in kT rather than T. The text version stores 293.6 K as 2.5300e-8 MeV, whereas the binary form stores it as 2.530049e-8 MeV. These are essentially rounding errors that do not impact the correctness of physics. Hence, the deviations do not impact the physical accuracy of the port. Therefore, these were accepted. Therefore, as far as ACE file parity is concerned, the port is satisfactory with respect to upstream NJOY2016.

With regard to the success of the Monte Carlo port, routes 1--5 were statistically indistinguishable from each other and within both 2 standard deviations and 50 pcm of each other. This number of 50 pcm was chosen as a suitable compromise between computational speed and accuracy. Even route 4, which bypasses ACE file generation, was negligibly different from route 1 within these bounds. The only exception was HST009, where the difference was 52 pcm from route 1, just barely over the limit, but still within two standard deviations. This extra 2 pcm was judged as sufficient as the statistics could always be improved later on. For the purposes of parity with upstream, this was considered sufficient. In fact, testing for route 4 could be done in the browser, as route 4 is the route taken for browser demos. This would satisfy both acceptance and qualification testing, and contribute to code documentation efforts.

Now, as to whether fission source convergence was achieved, web demos show that the Shannon entropy for all four cases converges within 40 inactive batches at 5000 particles per batch. For the five-route qualification calculations, LCT008 used the more conservative value of 10000 particles per batch. In each case, Shannon entropy reached an approximately stable plateau during the inactive generations and remained approximately stationary thereafter. This is shown in the figures:

\begin{figure}[H]
\centering
\includegraphics[width=\linewidth,height=0.75\textheight,keepaspectratio]{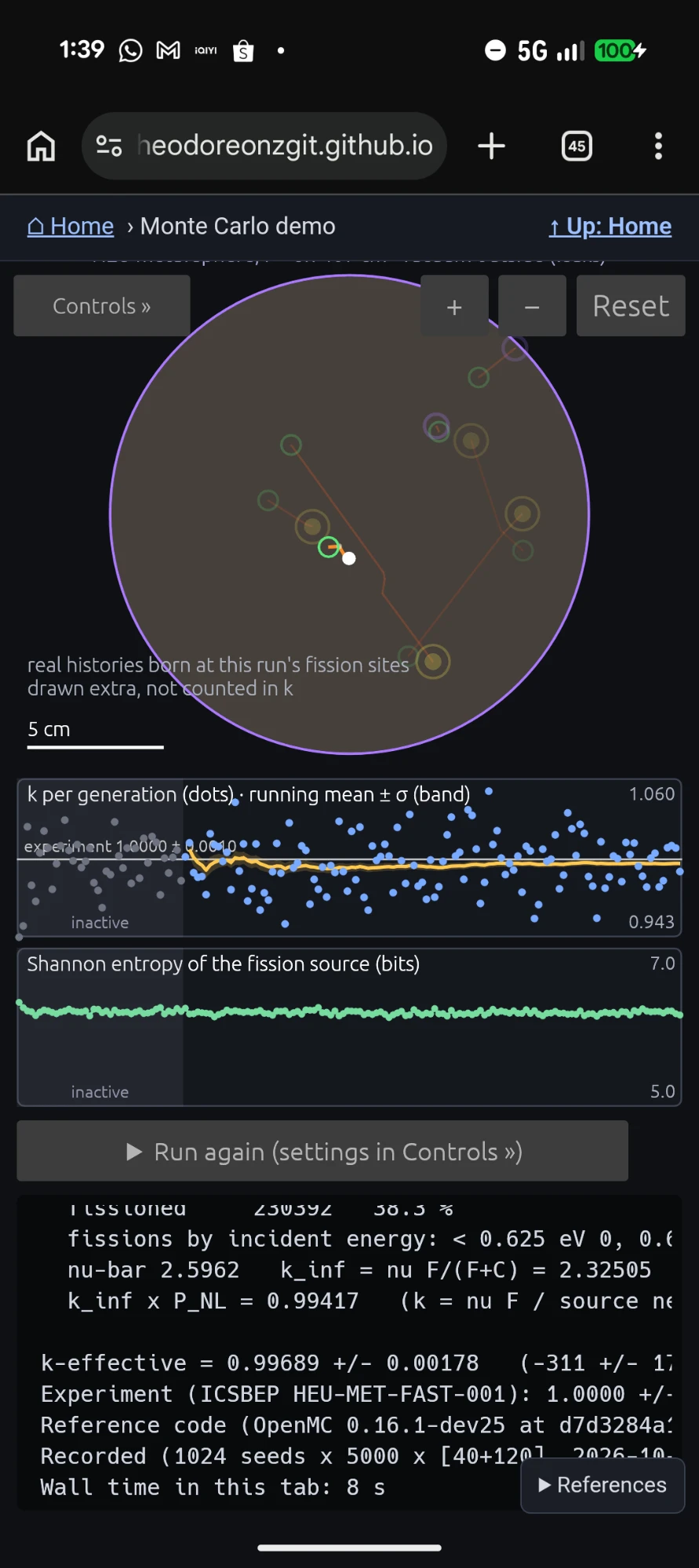}
\caption{Godiva (HEU-MET-FAST-001) run in the OutramMC web demo on an Android phone (5000 neutrons per batch, 40 inactive and 120 active batches, seed 1), showing $k$ per generation and the Shannon entropy of the fission source per generation, with the inactive generations shaded.}
\label{fig:godiva_phone_5000x40x120}
\end{figure}

\begin{figure}[H]
\centering
\includegraphics[width=\linewidth,height=0.75\textheight,keepaspectratio]{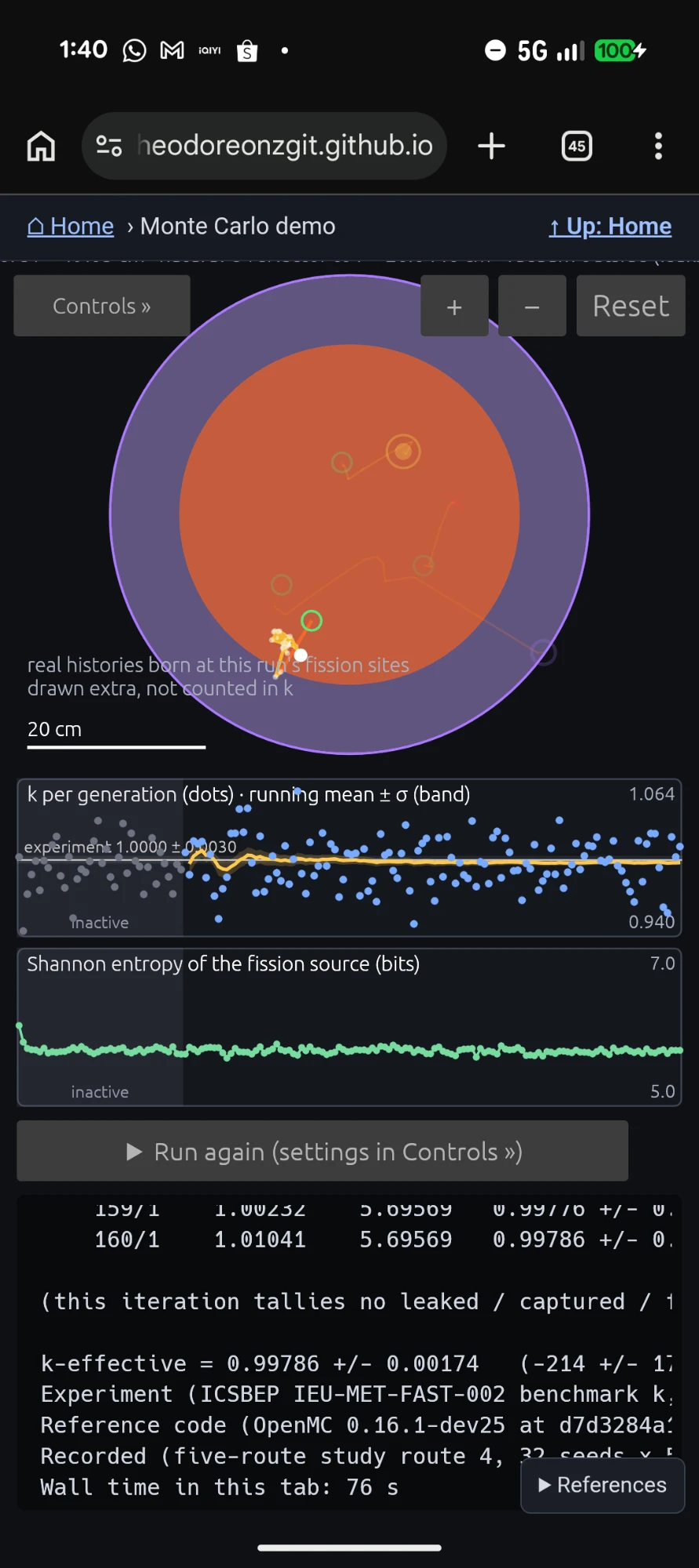}
\caption{Jemima (IEU-MET-FAST-002) run in the OutramMC web demo on an Android phone (5000 neutrons per batch, 40 inactive and 120 active batches, seed 1), showing $k$ per generation and the Shannon entropy of the fission source per generation, with the inactive generations shaded.}
\label{fig:jemima_phone_5000x40x120}
\end{figure}

\begin{figure}[H]
\centering
\includegraphics[width=\linewidth,height=0.75\textheight,keepaspectratio]{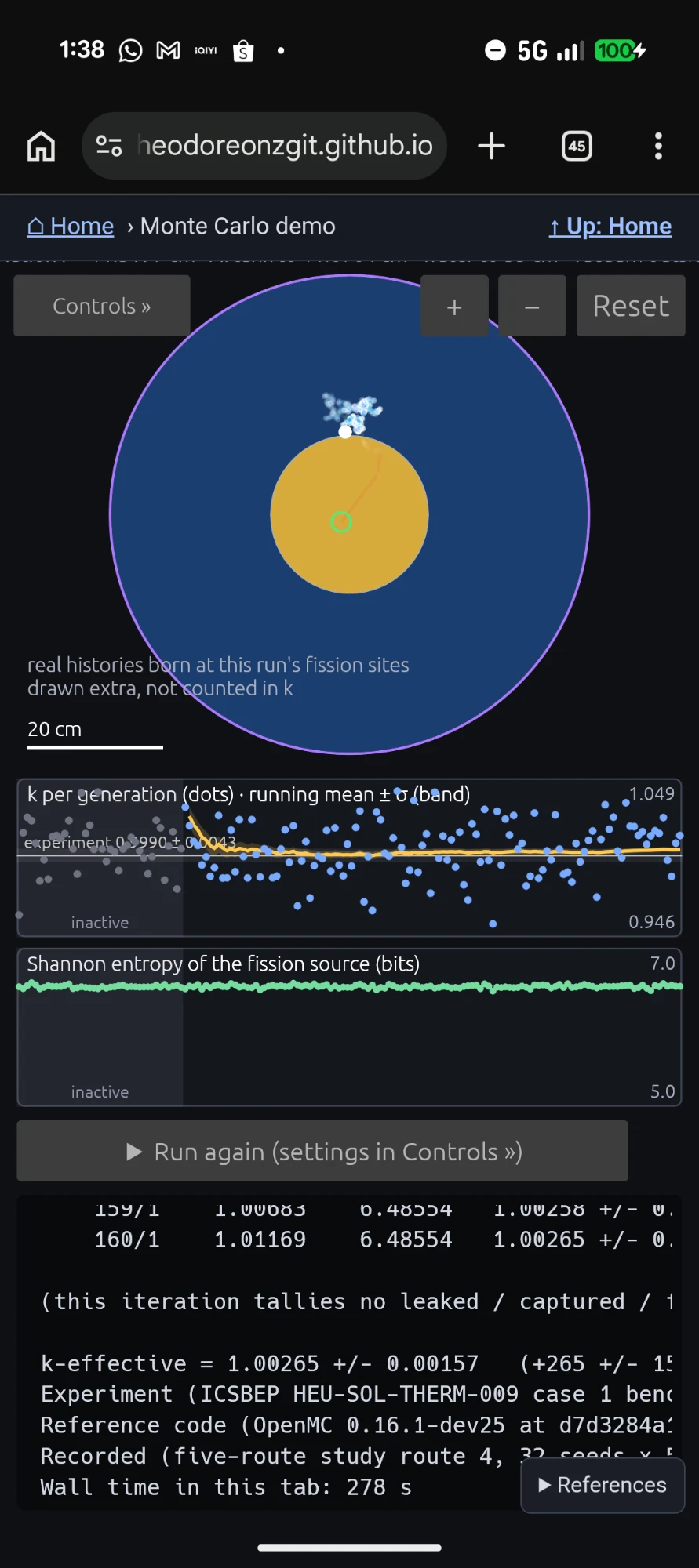}
\caption{HST-009 case 1 (HEU-SOL-THERM-009) run in the OutramMC web demo on an Android phone (5000 neutrons per batch, 40 inactive and 120 active batches, seed 1), showing $k$ per generation and the Shannon entropy of the fission source per generation, with the inactive generations shaded.}
\label{fig:hst009_phone_5000x40x120}
\end{figure}

\begin{figure}[H]
\centering
\includegraphics[width=\linewidth,height=0.75\textheight,keepaspectratio]{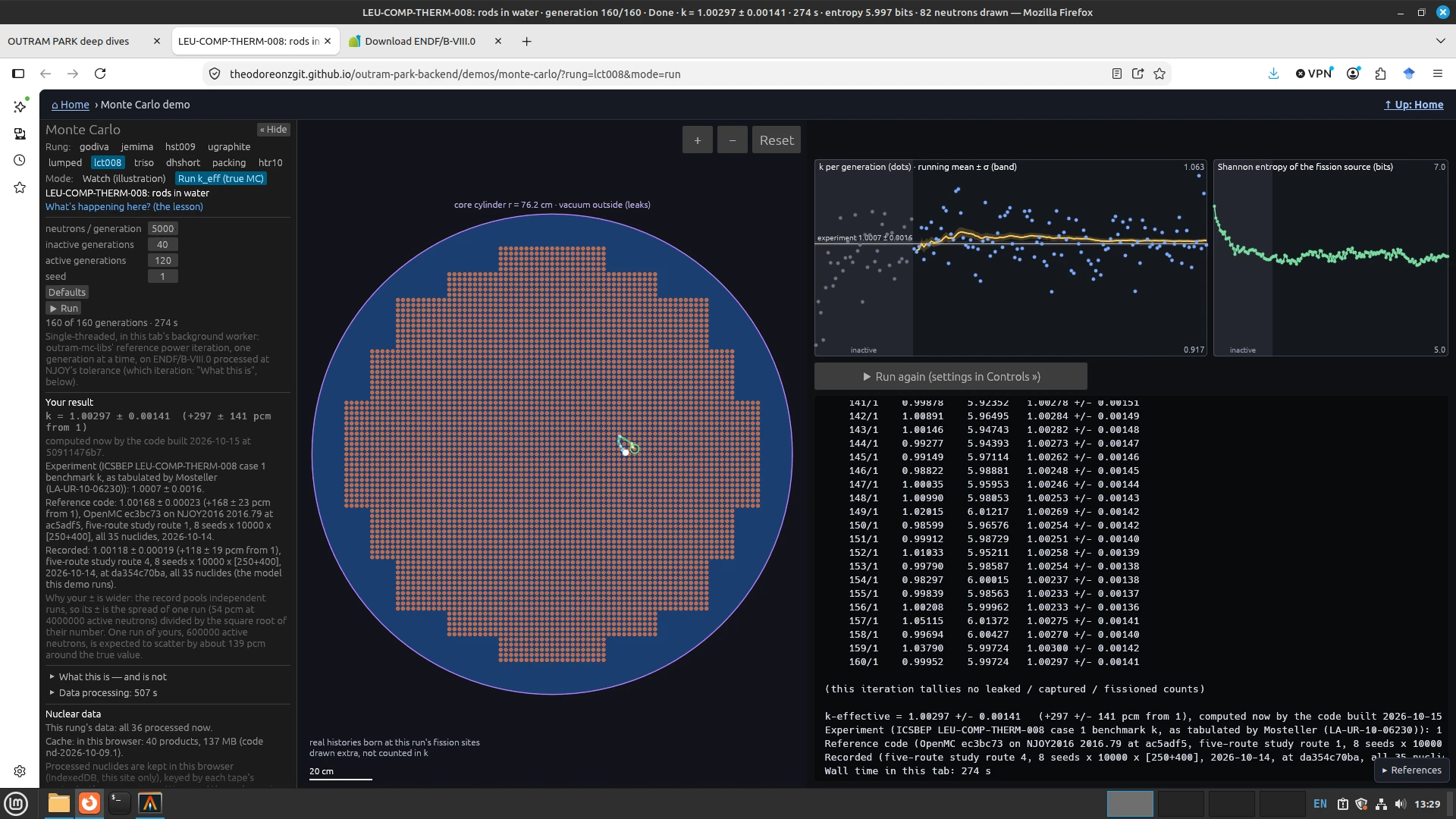}
\caption{LCT-008 case 1 (LEU-COMP-THERM-008) run in the OutramMC web demo in Firefox on a Linux desktop (5000 neutrons per batch, 40 inactive and 120 active batches, seed 1), showing $k$ per generation and the Shannon entropy of the fission source per generation, with the inactive generations shaded. The Shannon entropy converges within the 40 inactive batches at 5000 neutrons per batch.}
\label{fig:lct008_desktop_5000x40x120}
\end{figure}

\begin{figure}[H]
\centering
\includegraphics[width=\linewidth,height=0.75\textheight,keepaspectratio]{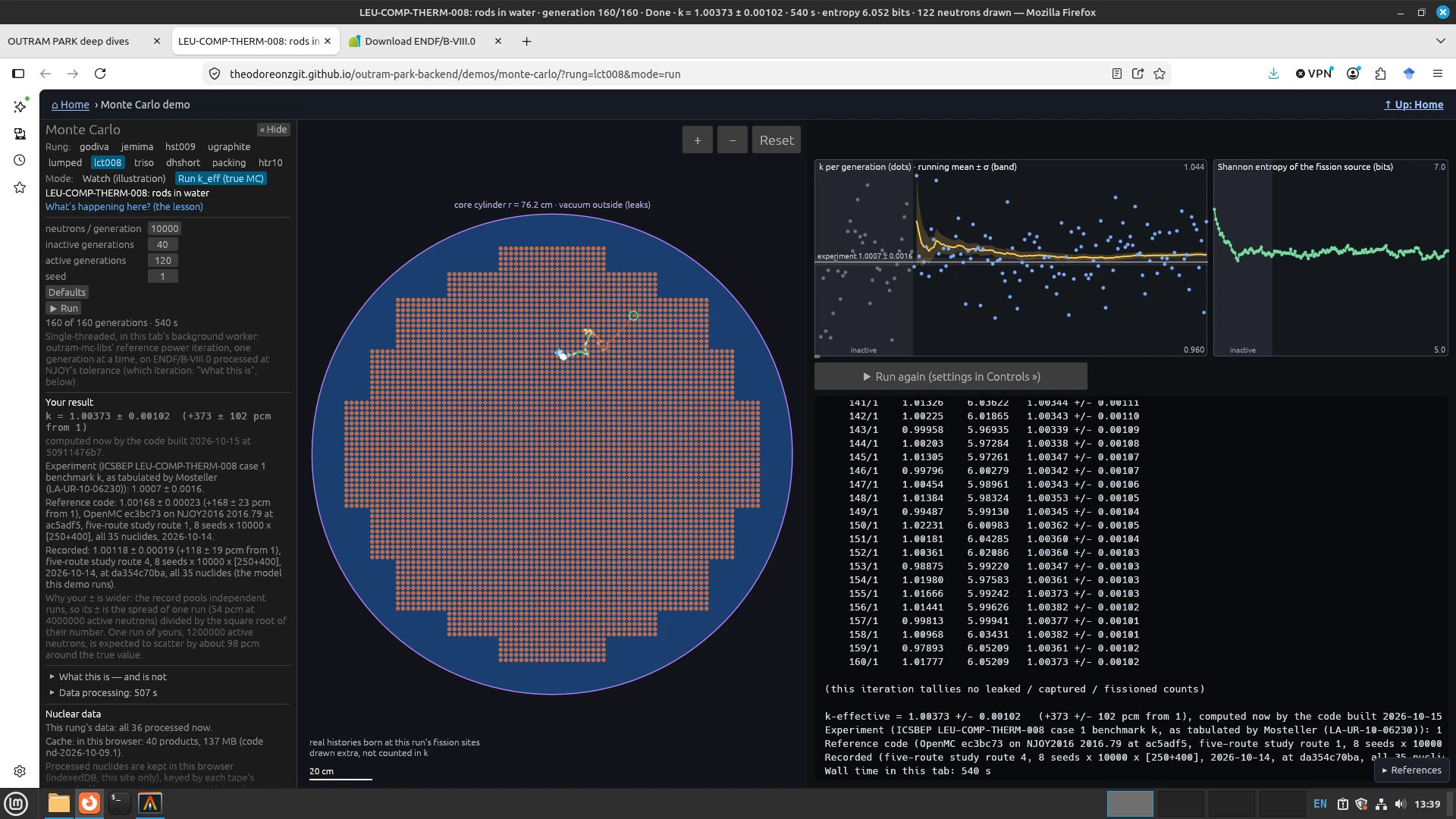}
\caption{LCT-008 case 1 (LEU-COMP-THERM-008) run in the OutramMC web demo in Firefox on a Linux desktop (10000 neutrons per batch, 40 inactive and 120 active batches, seed 1), showing $k$ per generation and the Shannon entropy of the fission source per generation, with the inactive generations shaded. The Shannon entropy converges within the 40 inactive batches at 10000 neutrons per batch.}
\label{fig:lct008_desktop_10000x40x120}
\end{figure}

These web demos were a convenient means to show Shannon entropy convergence after 40 batches, as not only did they not require compilation and preparation of nuclear data, the Monte Carlo settings could be changed easily using a slider. Therefore, runs at higher statistics could be run on both desktop PCs and mobile phones. This allowed the first author to make quick checks for the k-eigenvalue simulations as the Rust code used in the route runs was exactly the same as that deployed on web, only that the latter was compiled to WebAssembly.

Now, as the web demos and 32-seed testing show, code-to-code verification was done to within 2 standard deviations for routes 2--5. For validation, it is worth considering that only Godiva, HST009 and LCT008 were experimentally validated. This is still acceptable because the models, as coded in OpenMC and NJOY2016 in route 1, do not pass the Jemima benchmark either.

\begingroup\small
\setlength{\tabcolsep}{3pt}
\begin{longtable}{>{\raggedright\arraybackslash}p{0.233\linewidth}>{\raggedleft\arraybackslash}p{0.233\linewidth}>{\raggedleft\arraybackslash}p{0.233\linewidth}>{\raggedleft\arraybackslash}p{0.233\linewidth}}
\toprule
\textbf{Case} & \textbf{Benchmark \(k_\mathrm{eff}\)} & \textbf{Route 1 \(k_\mathrm{eff} \pm \mathrm{SEM}\)} & \textbf{$\Delta$ vs benchmark [pcm]} \\
\midrule
\endhead
Godiva (HEU-MET-FAST-001) & \(1.0000 \pm 0.0010\) & \(1.00016 \pm 0.00021\) & \(+16 \pm 21\) \\
Jemima (IEU-MET-FAST-002) & \(1.0000 \pm 0.0030\) & \(0.99594 \pm 0.00017\) & \(-406 \pm 17\) \\
HEU-SOL-THERM-009 case 1 & \(0.9990 \pm 0.0043\) & \(1.00124 \pm 0.00029\) & \(+224 \pm 29\) \\
LEU-COMP-THERM-008 case 1 (full 35 nuclides) & \(1.0007 \pm 0.0016\) & \(1.00161 \pm 0.00013\) & \(+91 \pm 13\) \\
\bottomrule
\end{longtable}
\endgroup

For purposes of translation, code-to-code verification is of a higher priority than validation as it shows that the functions were accurately ported over. For this reason, Jemima results were still acceptable and encouraging even if they did not meet the validation benchmark. It is likely that the model used was too simplified to replicate experimental results as both OpenMC and OutramMC produced similar results. Therefore, this was not a port defect. The reasons as to why the models were outside the band are left for future work because agentic porting is the main topic of this paper rather than validation.

However, validation gives evidence that the code can reproduce experimental data to within uncertainty bounds, and it still gives confidence that agentic porting was successful. For this purpose, Rust NJOY ports and OutramMC were validated against Godiva, HST009 and LCT008 to within experimental uncertainty. This completes qualification testing, and at this stage, the physics implementation of the code has no further reason to evolve as far as these benchmarks are concerned. This signals the agentic port's success and increased readiness for human review.

\subsection{Acceptance Testing}

For acceptance testing, the web demos show that Outram Park's fork of NJOY and OutramMC work well in the browser, across the first author's Google Pixel and Linux desktop devices. Moreover, the Godiva runs were performed at higher statistics on the author's colleague's Windows computer, thus showing its cross-platform compatibility on desktop.

\begin{figure}[H]
\centering
\includegraphics[width=\linewidth,height=0.75\textheight,keepaspectratio]{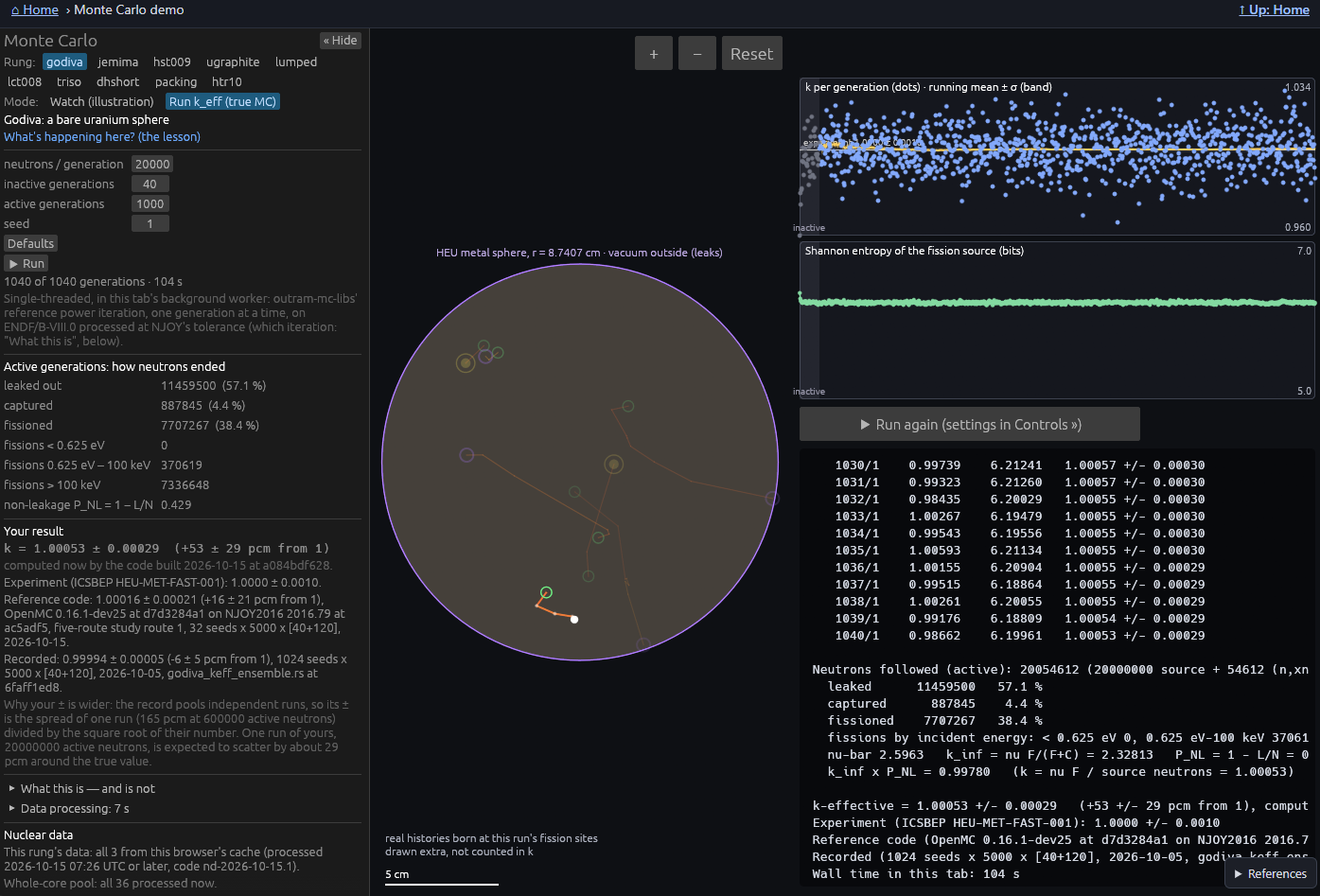}
\caption{Godiva on Windows browser, 20000 neutrons per cycle, 40 inactive cycles, 1000 active cycles.}
\label{fig:godiva_windows_20000x40x1000}
\end{figure}

In this particular run, the first author wanted to increase the statistics for the benchmark runs to 50000 particles per cycle and 1500 active cycles post Shannon entropy convergence, rather than the 32-seed statistics to replicate methods used in literature. The author and his colleague found that the web browser demo only allowed for 20000 particles and 1000 active cycles. This limitation was quickly addressed by feeding it back to AI agents to correct the defects. Once defects were corrected, HST009 was run at these same statistics on an iPad with an M2 chip. Not only did it demonstrate its capabilities to run on the iOS ecosystem, this correction also allowed for runs to be done at higher statistics on mobile devices to match methods used in literature.

\begin{figure}[H]
\centering
\includegraphics[width=\linewidth,height=0.75\textheight,keepaspectratio]{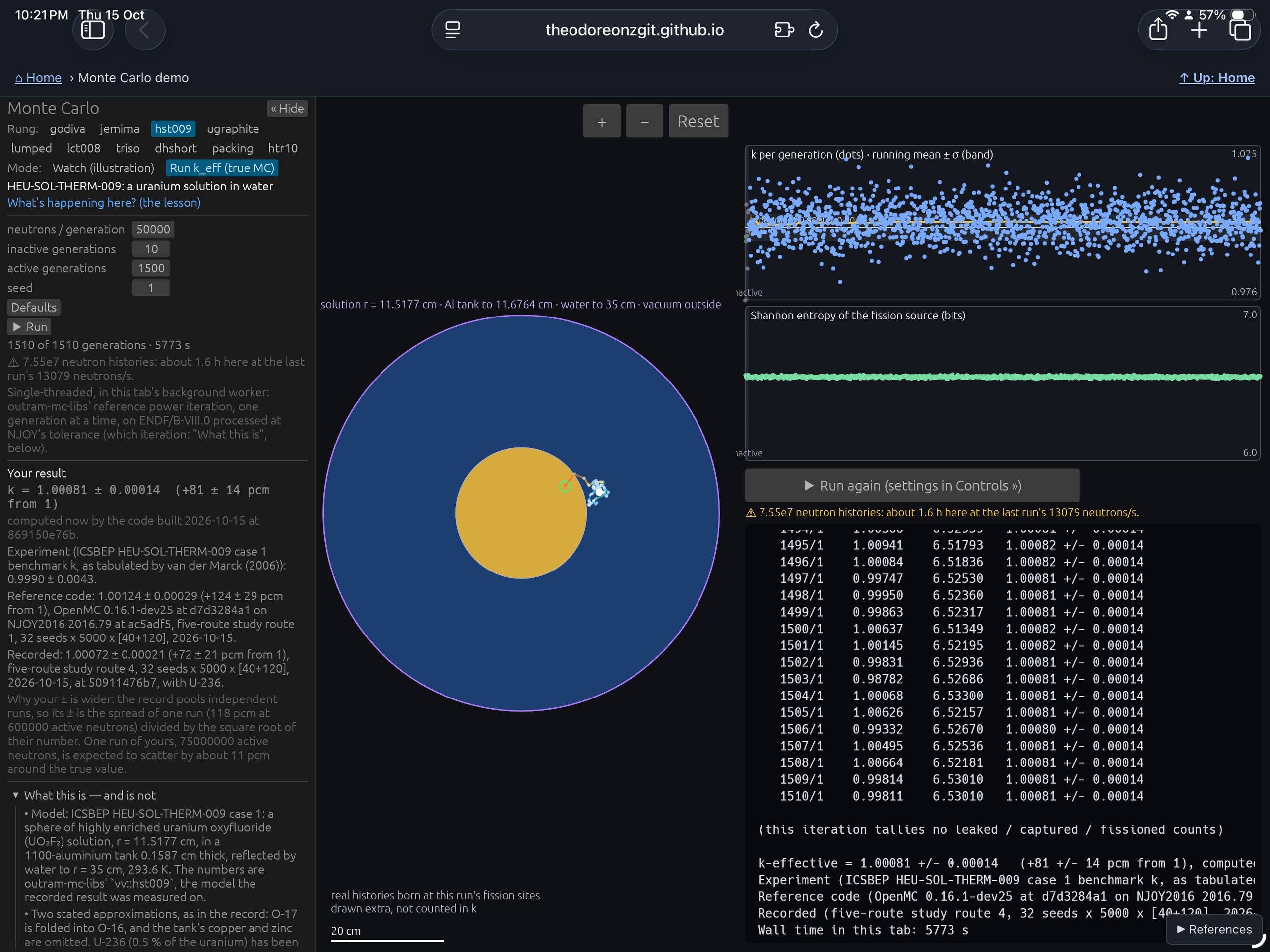}
\caption{High-statistics HST009 browser run on an M2 iPad (50000 neutrons per batch, 10 inactive and 1500 active batches): completed run, final generations.}
\label{fig:hst009_ipad_m2_50000x10x1500_final_end}
\end{figure}

\begin{figure}[H]
\centering
\includegraphics[width=\linewidth,height=0.75\textheight,keepaspectratio]{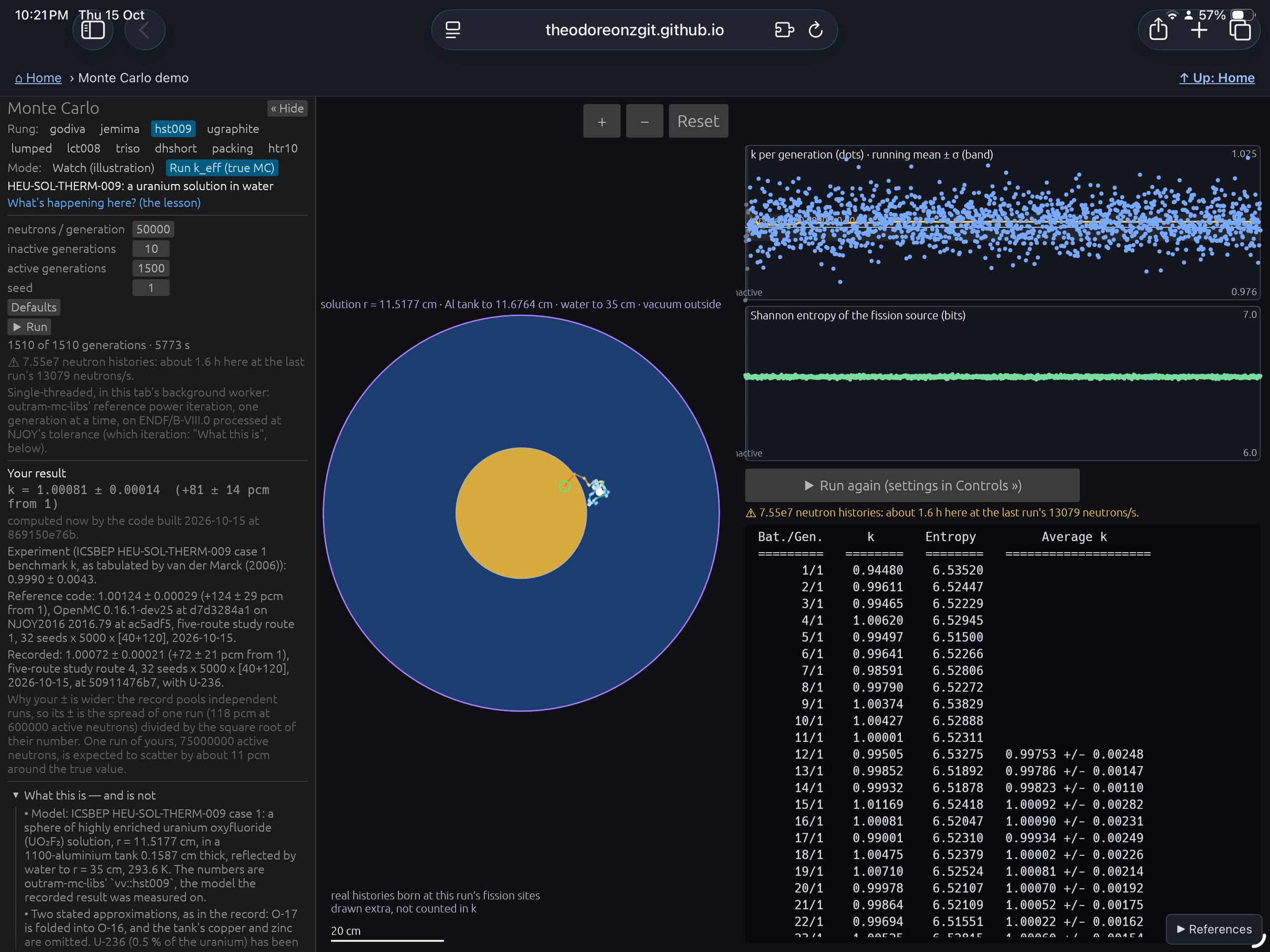}
\caption{High-statistics HST009 browser run on an M2 iPad (50000 neutrons per batch, 10 inactive and 1500 active batches): completed run, first generations.}
\label{fig:hst009_ipad_m2_50000x10x1500_final_start}
\end{figure}

On this iPad with an M2 chip, the HST009 runs took about 80--90 minutes.

The same LCT008 case, with the same settings and seed, was run in the Firefox browser on the first author's Arch Linux laptop. Here the demo ran single-threaded in one background worker, and the run took 11949 s, about 3.3 hours. It gave \(k_\mathrm{eff} = 1.00168 \pm 0.00014\), the same value, to the digits displayed, as the MacBook Air run, which used 7 workers in a different browser on a different operating system and processor.

\begin{figure}[H]
\centering
\includegraphics[width=\linewidth,height=0.75\textheight,keepaspectratio]{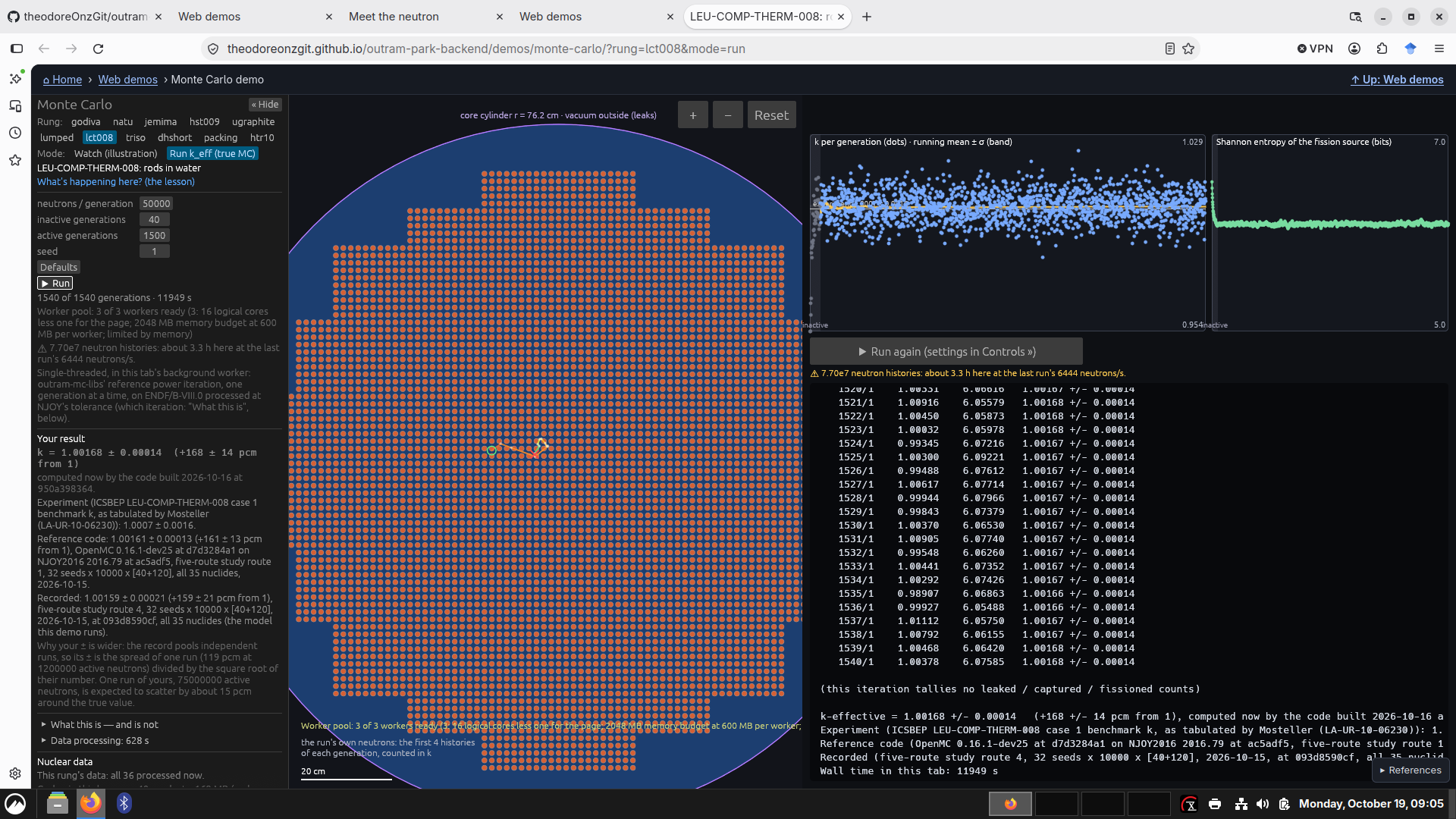}
\caption{High-statistics LCT008 browser run on Arch Linux: completed run, 50000 neutrons per batch, 40 inactive and 1500 active batches.}
\label{fig:lct008_arch_linux_50000x40x1500}
\end{figure}

The author also found out from colleagues that the LCT008 web demo ran on single threads, making it unnecessarily slow. This single-threaded case was run on Arch Linux, which took more than 3 hours to complete. This feedback was given to Claude agents, which then enabled multiple JavaScript workers to run LCT008 on multiple threads. However, this still provided useful data as it gave an oracle value for \(k_\mathrm{eff}\): \(k_\mathrm{eff} = 1.00168 \pm 0.00014\). This oracle value was important for comparison to ensure that multithreaded runs produced the same value. After the multithreaded implementation, LCT008 was also run in the Chrome browser on a MacBook Air (Apple M1, 2020, 16 GB, macOS 26.0), at 50000 particles per batch with 40 inactive and 1500 active batches. The run completed all 1540 batches in 3517 s, about 59 minutes, using 7 browser workers, and gave \(k_\mathrm{eff} = 1.00168 \pm 0.00014\). This is \(+98\) pcm from the experimental value of \(1.0007 \pm 0.0016\), and \(+7 \pm 19\) pcm from the route 1 reference of \(1.00161 \pm 0.00013\). This showed that the multithreaded run was not only faster, but was able to reproduce the single-threaded result deterministically. This provided evidence that multithreading was performed without compromising accuracy.

\begin{figure}[H]
\centering
\includegraphics[width=\linewidth,height=0.75\textheight,keepaspectratio]{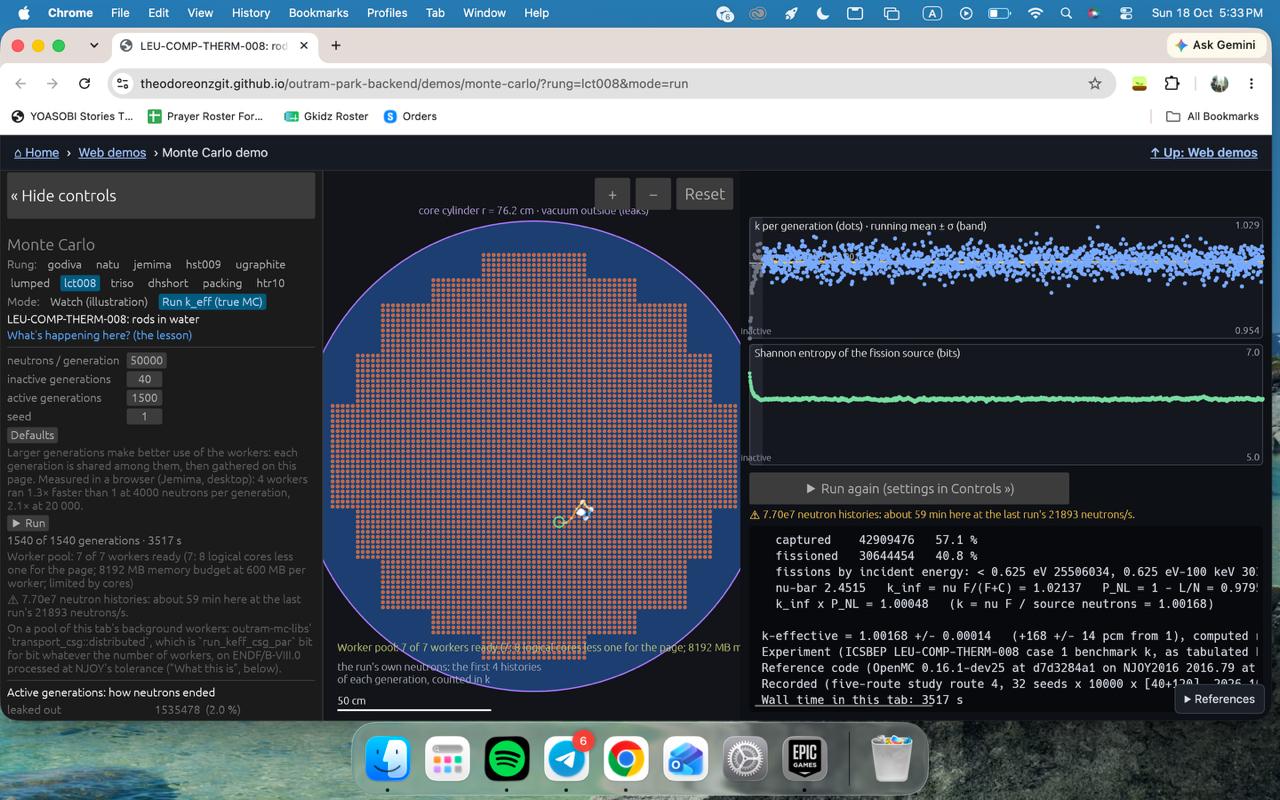}
\caption{High-statistics LCT008 browser run on an M1 MacBook Air: completed run, 50000 neutrons per batch, 40 inactive and 1500 active batches.}
\label{fig:lct008_macbook_air_m1_50000x40x1500}
\end{figure}

Godiva was also run at 50000 particles per batch, with 40 inactive and 1500 active batches, in the browser on the first author's Android phone. The run took 866 s, about 15 minutes, and gave \(k_\mathrm{eff} = 0.99996 \pm 0.00015\), or \(-4 \pm 15\) pcm from the experimental value of 1.0000.

\begin{figure}[H]
\centering
\includegraphics[width=\linewidth,height=0.75\textheight,keepaspectratio]{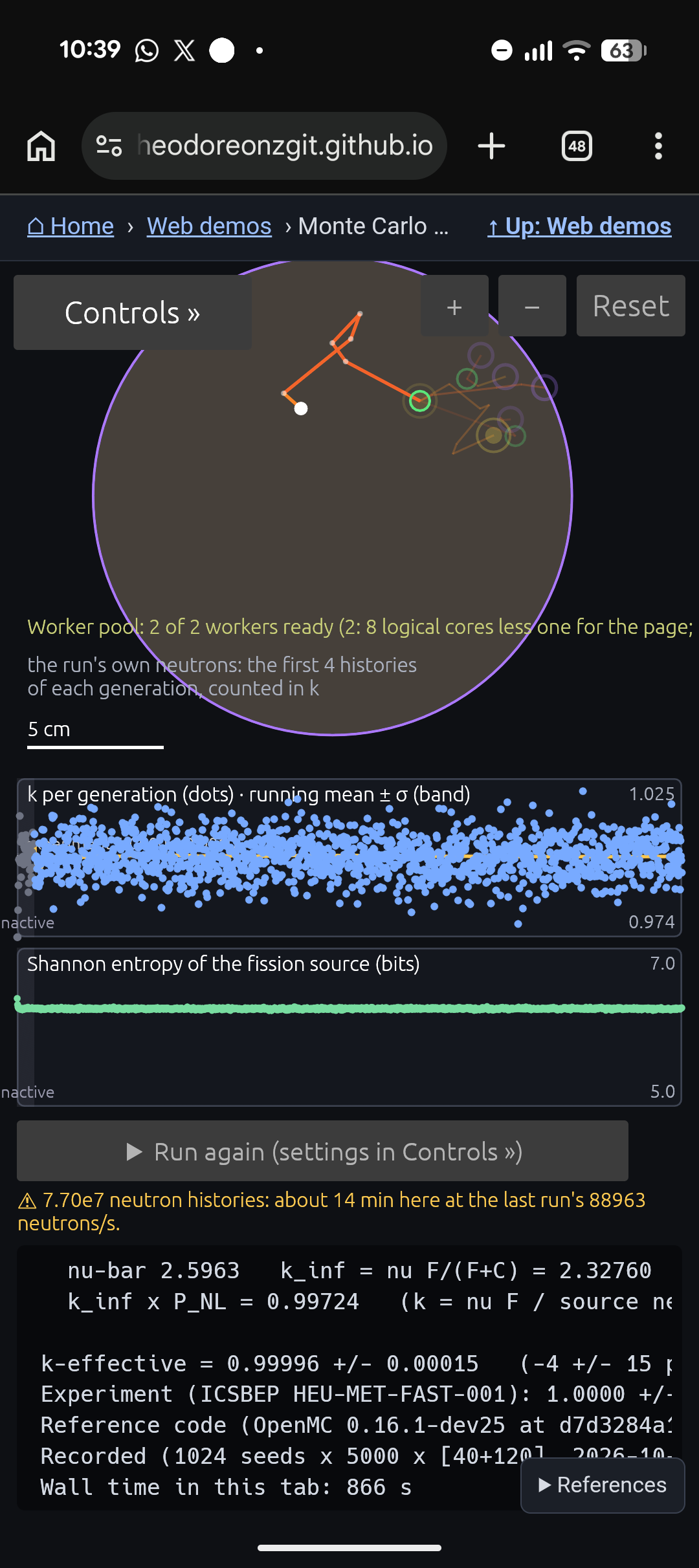}
\caption{High-statistics Godiva browser run on an Android phone: completed run, 50000 neutrons per batch, 40 inactive and 1500 active batches.}
\label{fig:godiva_phone_50000x40x1500}
\end{figure}

Thus, acceptance testing here demonstrated the portability of Outram Park's NJOY fork and OutramMC web demos across desktop Windows, Linux and macOS, as well as Android and iOS platforms. Of course, one does not need to have a collection of devices for acceptance testing. Commonly accepted software practices include the use of GitHub Actions for continuous integration so that tests are done across all targeted platforms. This includes macOS in addition to the aforementioned platforms used. GitHub Actions allows for automated publishing of web demos and documentation to GitHub Pages as well. This was agentically achieved and made acceptance testing and document generation a relatively low-effort endeavour. Moreover, these tests ran each of the three validation cases at higher statistics, according to Brown's recommendations \cite{brown2009mcreview} and Gao's settings on the order of 50000 neutrons per cycle and 1500 active cycles after Shannon entropy convergence \cite{gao2013rmc}.

Of note in cross-platform testing is that the numerical libraries used in different platforms have slight differences, resulting in rounding errors that differ in the last ulp when power functions are involved. Therefore, rather than relying on numerical libraries written in C for each platform, a pure Rust translation of the GNU Scientific Library (GSL) \cite{galassi2002gnu} was performed and made to be no\_std with alloc. This allows Chebyshev polynomials used in NJOY as well as exponential and power functions to be consistent across all platforms, thus ensuring that even these rounding errors were eliminated. This crate was agentically verified against upstream GSL and other C libraries it was translated from, and was named PETIR, after the Petir LRT station in Singapore. PETIR stands for Polynomials, Equations, Transforms, Integration and Roots.

The publication of web demos and use of CI meant that the code could be deployed cross-platform and yield numerically accurate and consistent results. This ensured that there was no further motivation for code to be agentically patched for the purposes of cross-platform compatibility. Thus, it stabilises the code even further, signalling even more readiness for human review.

Web browser demos also allow the end user to reproduce results easily. Perhaps not for the 32-seed versions (as one has to manually key in 32 seeds to run), but for a high-statistics run version where a high number of neutrons and hundreds of batches are used. This allows for easy reproducibility of aforementioned results. The web demo screenshots show explicitly what seed was used, and the exact settings used for the simulation. Given that these are pseudorandom number generators, results are deterministically reproducible. Moreover, there is no requirement to compile code; one simply has to visit the website and run the simulations on the phone or laptop to repeat the aforementioned calculations. This means that results are substantially easier to reproduce than having to compile NJOY and OpenMC via terminal and run the cases.

\subsection{Documentation and Human Review}

Human review for code often starts at documentation. For instance, the first author looked at OpenMC documentation (v0.12.0) when first starting installation. Along the way, Bash and Windows Subsystem for Linux were learnt in order to run and install OpenMC. Then, the first author looked at tutorial cases in Jupyter notebooks in order to learn how to run OpenMC and use its Python API. The success of using OpenMC was largely due to its simple-to-use API, documentation and tutorial cases. Moreover, OpenMC also contained plenty of theory documentation, so that with sufficient time spent, one could understand its code and algorithms such as surface tracking and the track length estimator for tallies.

This shows the importance of having tutorial cases and code documentation in order to understand how to use the code and which physics concepts it implements. In learning from OpenMC's well-documented Python API, the same approach was used for OutramMC and the Outram Park fork of NJOY. In this manner, there should be ample code documentation explaining the theory of what scientific concepts the code implements. For the API documentation, however, it is specifically left out at this early stage because the API is still unstable and evolving as agents make rapid changes to the codebase. The scientific concepts, however, do not change, and serve as stable anchors for code documentation.

For this documentation process, the author's personal experience played a role in shaping the documentation. While the mathematical equations were important to understand, the first author found them difficult to absorb on the first read. It was easier for the author to have a visual understanding of the physics-based processes going on in code to then interest the first author in the mathematical expressions implemented in code. Therefore, web demos became a central piece of the code documentation itself, not just playing a role in acceptance testing and qualification testing, but also in acquainting would-be readers with visually appealing simulations to interest the reader to read and understand more of the physics concepts and the code itself. Such demos make the code easier to review as they help people new to the code acquaint themselves with the scientific concepts expressed in code, before going into the syntax.

Therefore, interactive web demos and documentation play an important role in preparing the code to be human reviewed. This is especially important where the human driver behind the AI agents has little to no experience in developing the code that the AI agents translated and developed. In this process, the AI agents were instructed to create web demos and documentation from the Outram Park codebase to slowly educate and acquaint the first author with the physics-based concepts. The end result is that by the time the author is done understanding and reviewing the code, the byproduct is that the code is already well documented and more auditable in comparison to code that the author hand coded without agents.

This was the experience when the author coded TUAS from scratch \cite{ong2024tuas}: when hand coding TUAS, the author understood the code thoroughly, but no one else did. Heat transfer and thermal resistance concepts in TUAS obvious to the author were not obvious to others. When the first author wished to share his code, people lost interest when the documentation was severely lacking. This was despite the fact that there were several documentation comments written in TUAS, in addition to comments left in code. The lesson learnt was that documentation was important, and the documentation has to be custom-made for a person who doesn't know much about the code, nor the concepts it implements. In this regard, documentation comments and comments in code were insufficient. The first author had to spend tedious amounts of effort to write the equations down in a wiki to let others know about the code. Not only that, the first author also had to spend extra effort to step into the shoes of a new user in order to write good documentation. This was sometimes not achievable as there were other priorities at work.

With agentically derived code, however, the author is, serendipitously, in the position of a new user, because the code was never written by the human driver. Documentation that the human driver of AI agents needs would be tailor-made for someone who does not know the code well. Thus the perspective taking step is no longer required. Moreover, when the human driver, trying to understand code, asks AI to generate code documentation, the documentation itself would be actively reviewed from the outset, thus ensuring that it is readable and useful for the end user. This process guarantees that the first draft of documentation produced has actually been read and used, with pain points given as feedback to the AI to regenerate the documentation. The end result, as mentioned, is human reviewed documentation that actually helps the end user and makes the code more auditable. In this regard, the process described here makes the code thoroughly documented from the outset. This then makes the code easier to review, thus indicating that it is more ready for review than before code documentation was agentically generated, whether in the form of web demos or prose.

\begin{figure}[H]
\centering
\includegraphics[width=\linewidth]{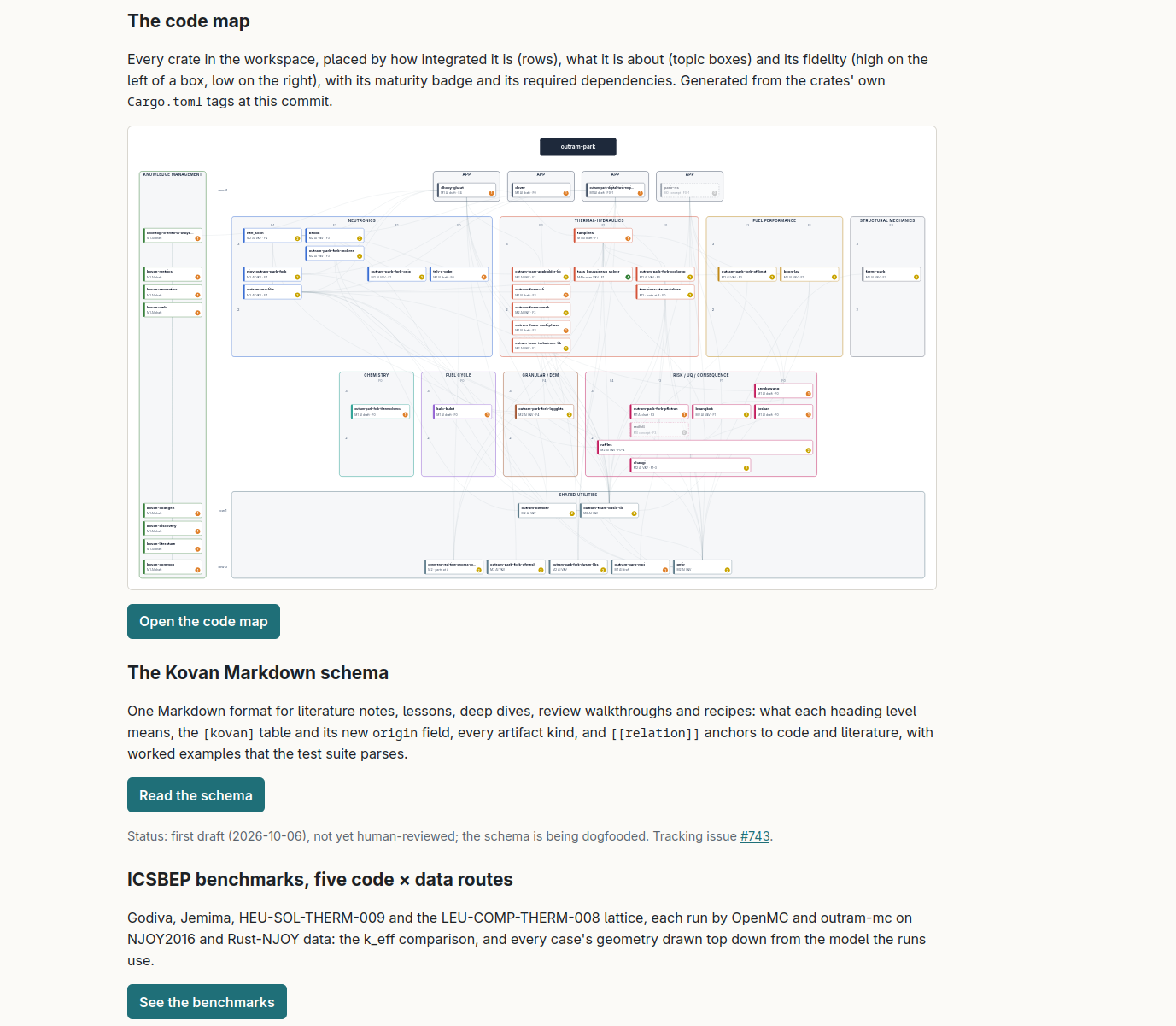}
\caption{The code map page, which was formerly the homepage of the first draft of the Outram Park documentation: every crate in the workspace placed by how integrated it is, its topic and its fidelity, followed by links to the Kovan Markdown schema and the ICSBEP benchmarks. Page generated agentically (Claude Code) under the first author's direction; screenshot by the first author, 19 October 2026.}
\label{fig:docs_first_draft_codemap}
\end{figure}

In this first draft of documentation, the documentation was found to be messy and confusing. The first author expected a lesson in physics first to understand the code, but was then thrown a series of tabs with which to understand code. This was the first pain point of the documentation. Therefore, the web documentation was agentically revised to have a simpler UI, one reminiscent of Google's search engine homepage, for user friendliness \cite{google2009homepage}.

\begin{figure}[H]
\centering
\includegraphics[width=\linewidth]{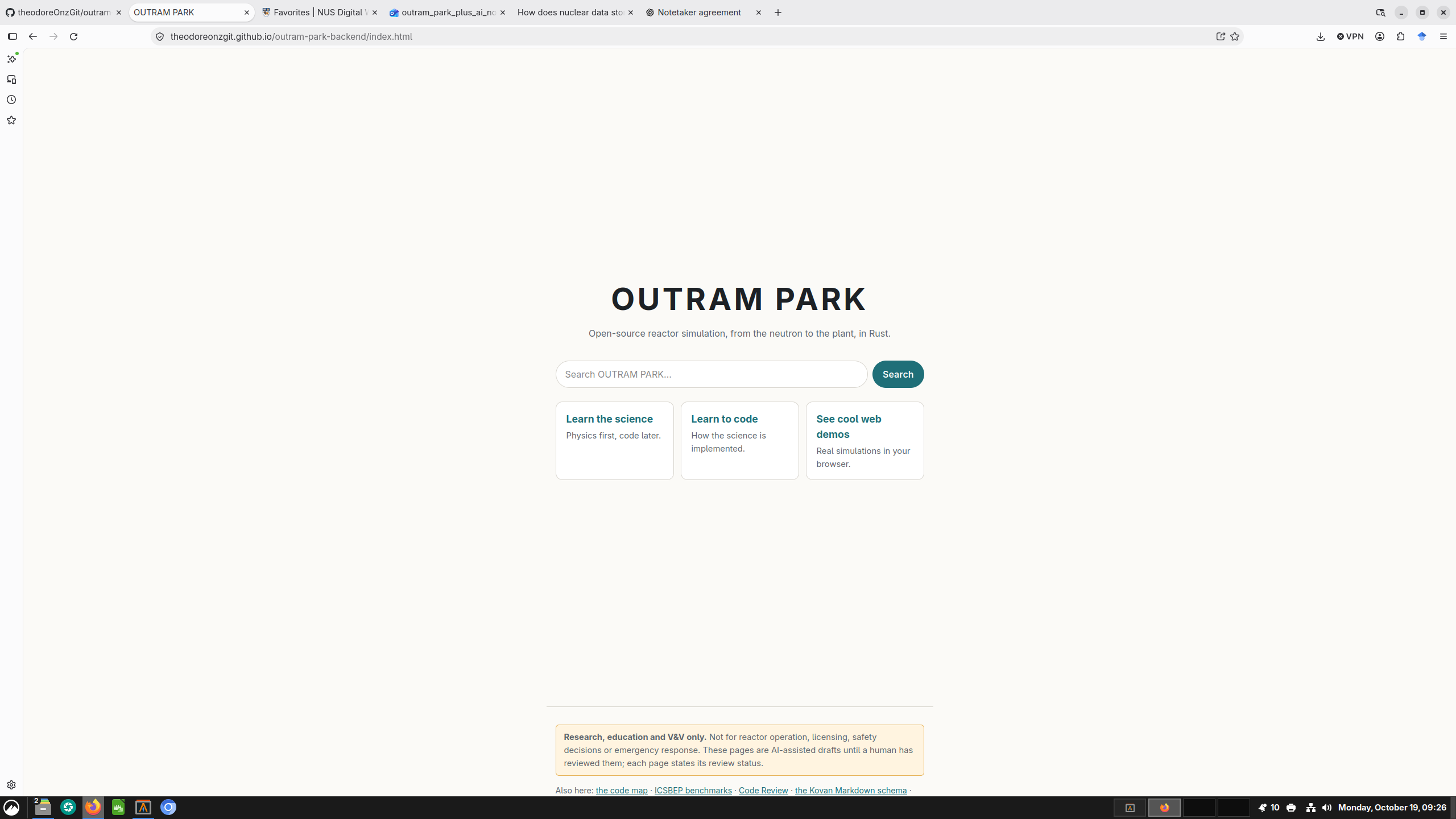}
\caption{Revised Outram Park web documentation homepage, showing the simplified search-engine-inspired interface. Page generated agentically (Claude Code) under the first author's direction; screenshot by the first author, 19 October 2026.}
\label{fig:docs_homepage}
\end{figure}

Aesthetics improves the overall usability of the site documentation, as research indicates that users often prefer websites with low complexity and high prototypicality \cite{tuch2012visualcomplexity}. Prototypicality refers to the degree to which websites follow certain prototypes, or mental models \cite{tuch2012visualcomplexity}. For instance, if one thinks about a search engine or homepage, one might immediately think about Google's search page. Therefore, in documentation navigation, a user friendly home search page should look like Google. This homepage facilitates giving a good first impression, often formed within 50 ms of seeing the page \cite{lindgaard2006attention}.

The documentation itself was then refactored agentically into a search bar resembling Google's search bar, and three options underneath targeting three specific target audiences. Firstly, those who want to learn science without code. Secondly, those who want to learn the code itself, and users of such documentation would then view visually appealing API documentation. Thirdly, those who want to view Outram Park's web demos, wherein both technical and non-technical users can see the aforementioned web demos in addition to other ones planned in future.

\begin{figure}[H]
\centering
\includegraphics[width=\linewidth]{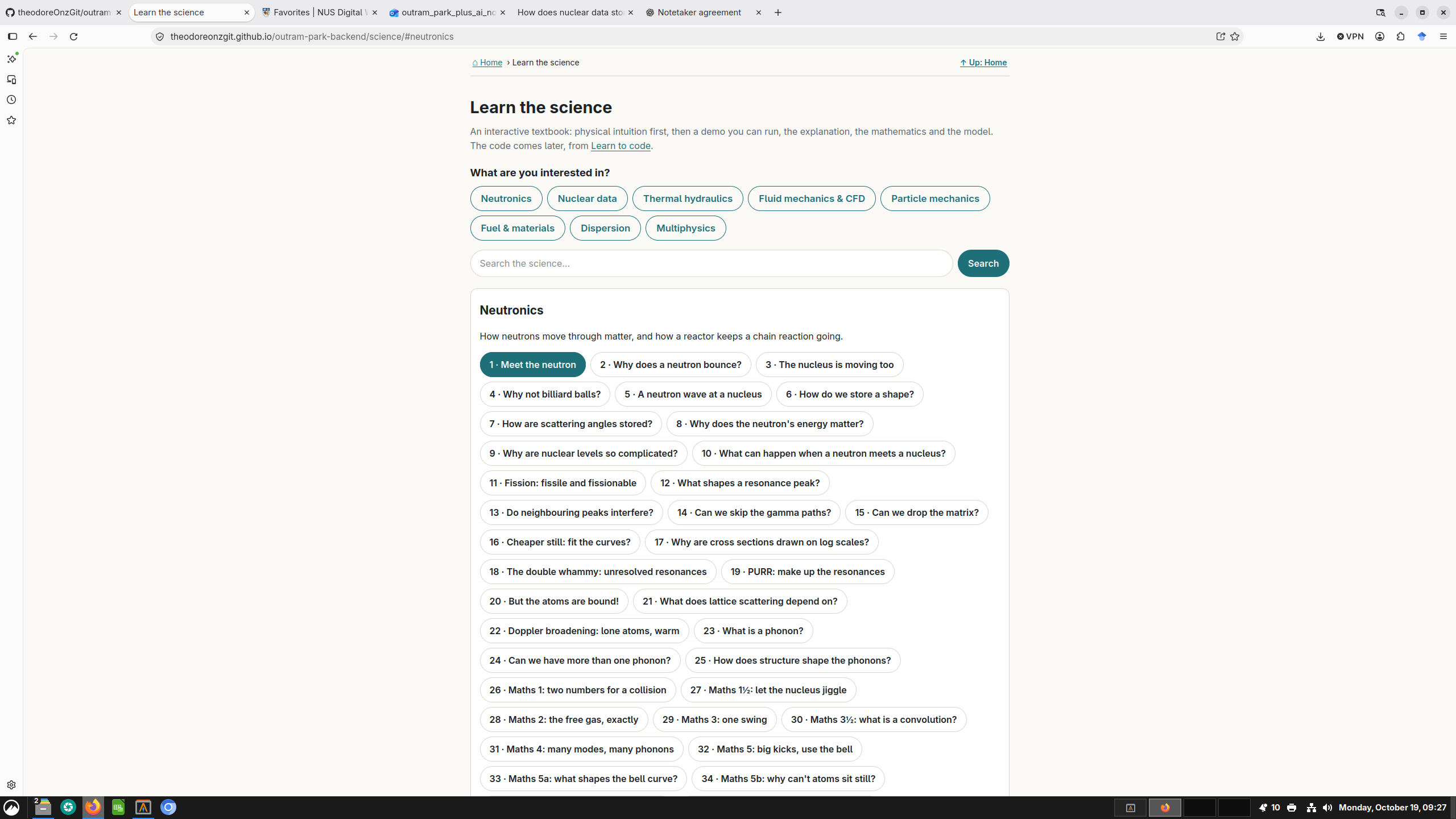}
\caption{The ``Learn the Science'' index: the topics, and the neutronics pages in reading order. Page generated agentically (Claude Code) under the first author's direction; screenshot by the first author, 19 October 2026.}
\label{fig:docs_learn_science}
\end{figure}

\begin{figure}[H]
\centering
\includegraphics[width=\linewidth]{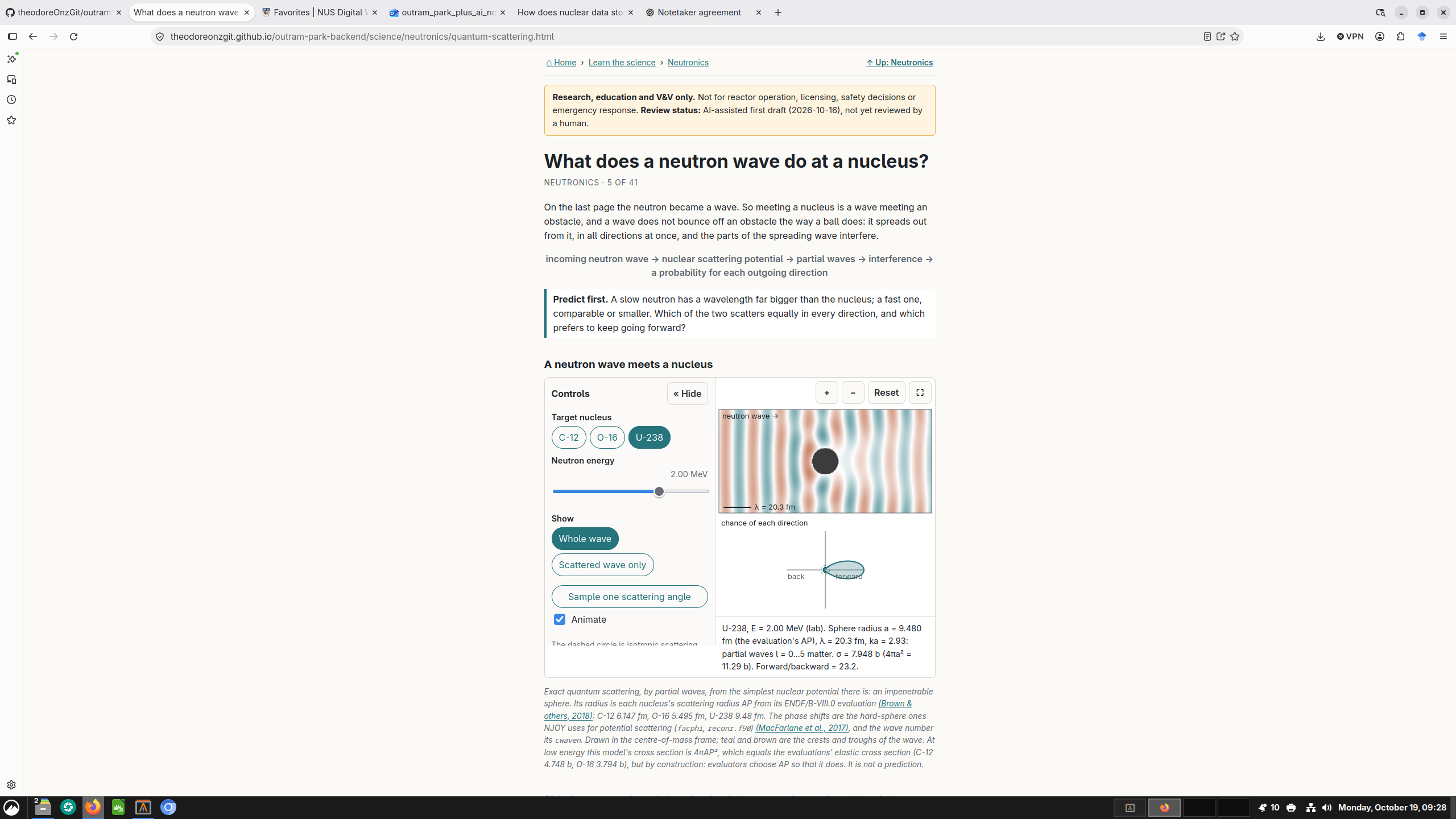}
\caption{Learn the Science page ``What does a neutron wave do at a nucleus?'', with its interactive wave-scattering demonstration. Page generated agentically (Claude Code) under the first author's direction; screenshot by the first author, 19 October 2026.}
\label{fig:docs_neutron_waves}
\end{figure}

\begin{figure}[H]
\centering
\includegraphics[width=\linewidth]{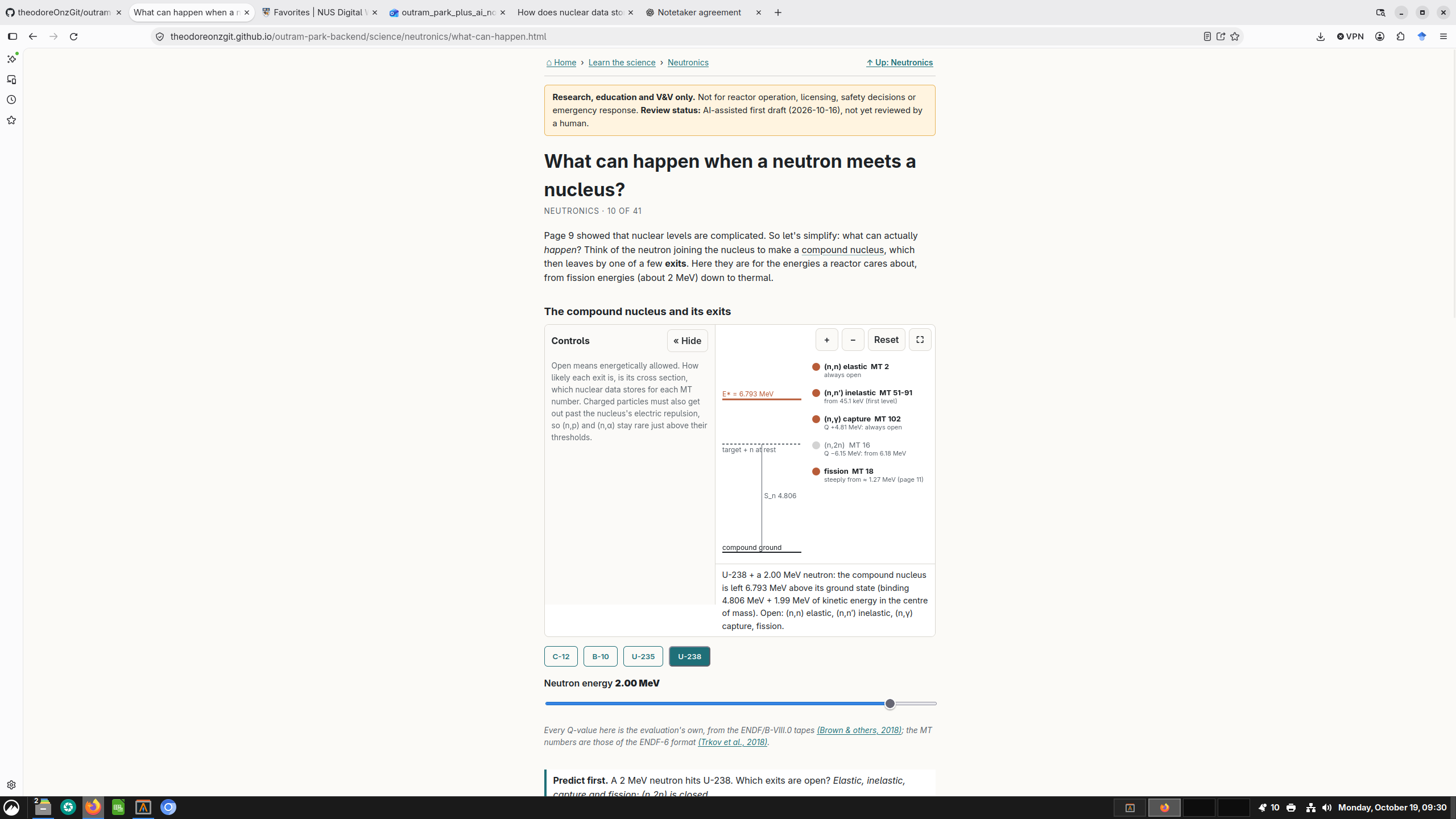}
\caption{Learn the Science page ``What can happen when a neutron meets a nucleus?'': the compound nucleus and its exits. Page generated agentically (Claude Code) under the first author's direction; screenshot by the first author, 19 October 2026.}
\label{fig:docs_compound_nucleus}
\end{figure}

\begin{figure}[H]
\centering
\includegraphics[width=\linewidth]{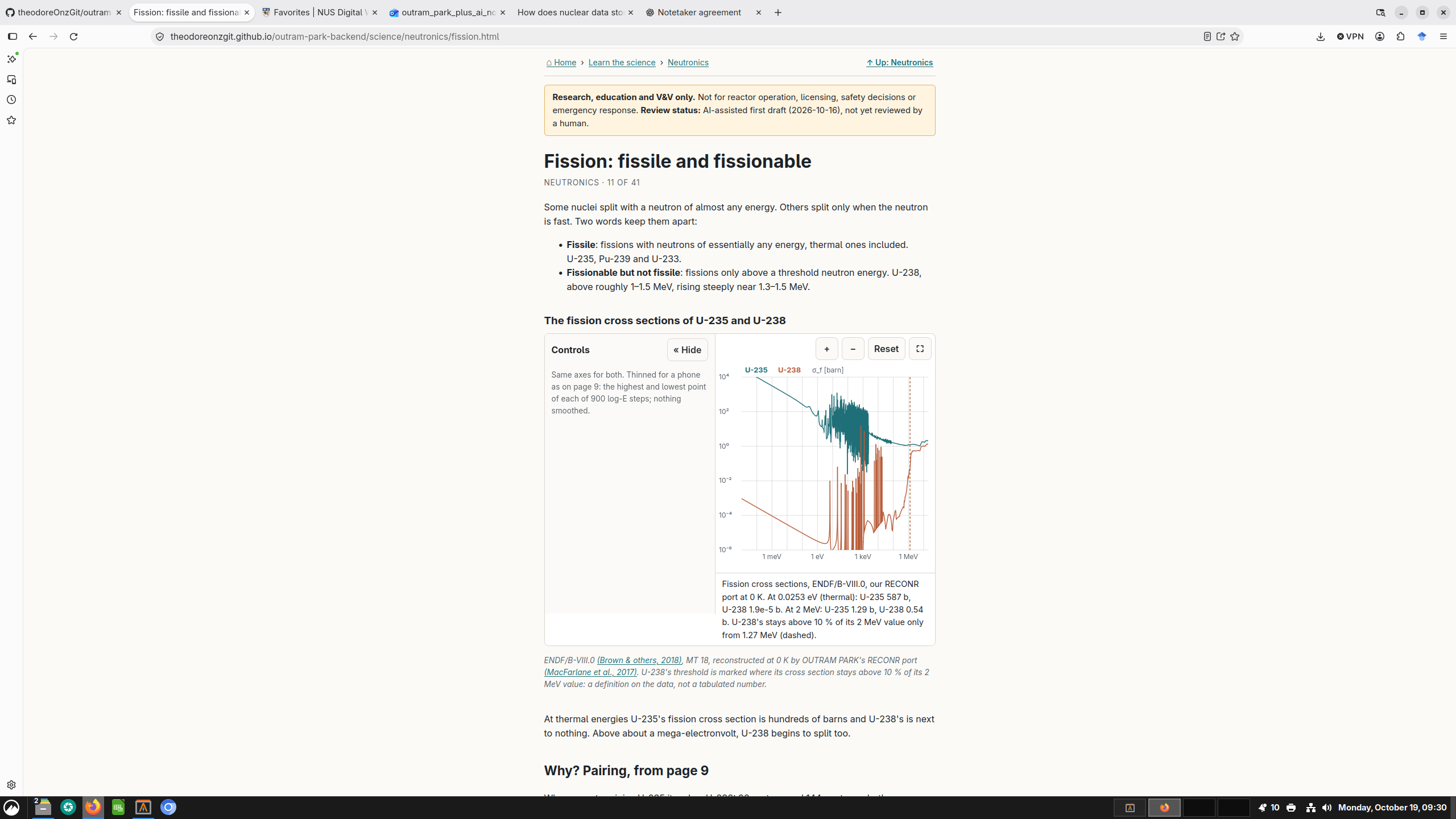}
\caption{Learn the Science page ``Fission: fissile and fissionable'', comparing the fission cross sections of U-235 and U-238. Page generated agentically (Claude Code) under the first author's direction; screenshot by the first author, 19 October 2026.}
\label{fig:docs_fissile}
\end{figure}

The ``Learn the Science'' docs contain concepts important in understanding code. In this second iteration, the first author prompted AI agents to first include nuclear physics background, as the author was familiar with it due to the basic nuclear engineering coursework taken in graduate studies. (Figures~\ref{fig:docs_neutron_waves}, \ref{fig:docs_compound_nucleus} and \ref{fig:docs_fissile}) These give visually appealing simulations to help readers visualise the concept of neutrons as matter waves scattering upon the nucleus, thus giving conceptual background to energy levels in the atom, resonances and the interference patterns resulting in non-isotropic scattering patterns.

From there on, the first author started learning about concepts from Claude Opus 5.5 agents from single resonance formalisms, which were easy to connect conceptually to using shape functions (versus interpolated data in PENDF) to represent a resonance. This gave rise to understanding the various means of representing resonance data in ENDF files, such as single-level Breit-Wigner resonances. The concepts of wave functions were then linked to interference and how multiple resonances were represented using the R-matrix. The first author then, understanding surrogate models, understood that simplifications aided in computation speed. Therefore, the simplifications of Reich-Moore and Multi-Level Breit-Wigner were documented agentically. Quoting the documentation, in Reich-Moore the gamma decay paths simplify the R-matrix by lumping multiple decay channels into ''\textbf{one total capture width $\Gamma_\gamma$ per level}, the eliminated width'' \href{https://theodoreonzgit.github.io/outram-park-backend/science/neutronics/reich-moore.html\#ref-njoy2016}{(MacFarlane et al., 2017)} and R-matrix Limited extends this concept to ``unlimited numbers and types of channels, charged-particle exits included'' \href{https://theodoreonzgit.github.io/outram-park-backend/science/neutronics/reich-moore.html\#ref-trkov2018endf6}{(Trkov et al., 2018)}. The Multi-Level Breit-Wigner (MLBW) resonance formalism then performs simple superposition for fission and capture resonance cross sections. The distinction here is that the cross section, which is a proportionality constant for interaction probability, is analogous to the wavefunction squared. The proper way to obtain probabilities is to sum up wavefunctions and then perform the squaring of summed up amplitudes. This proper interference treatment was done for elastic scattering, but the simple addition of cross sections was done for fission and capture. From the generated documentation, it ``allows multilevel computations only for elastic; for other types of partial cross sections the formalism is single-level'' \href{https://theodoreonzgit.github.io/outram-park-backend/science/neutronics/mlbw.html\#ref-trkov2018endf6}{(Trkov et al., 2018)}''.  These concepts, now in documentation and lesson format, then helped the author understand the jargon used in MacFarlane's NJOY manual and understand the resonance formalisms used in NJOY. The most extensive simplification was the Adler-Adler model, which emphasised data fitting more than the preceding models and is not as extensively used anymore according to agents. In constructing this documentation, the first author left sufficient breadcrumbs for himself to understand these resonance formalisms as well as for other users. This is crucial in understanding RECONR algorithms in NJOY and the Rust port discussed in this work.

\begin{figure}[H]
\centering
\includegraphics[width=\linewidth]{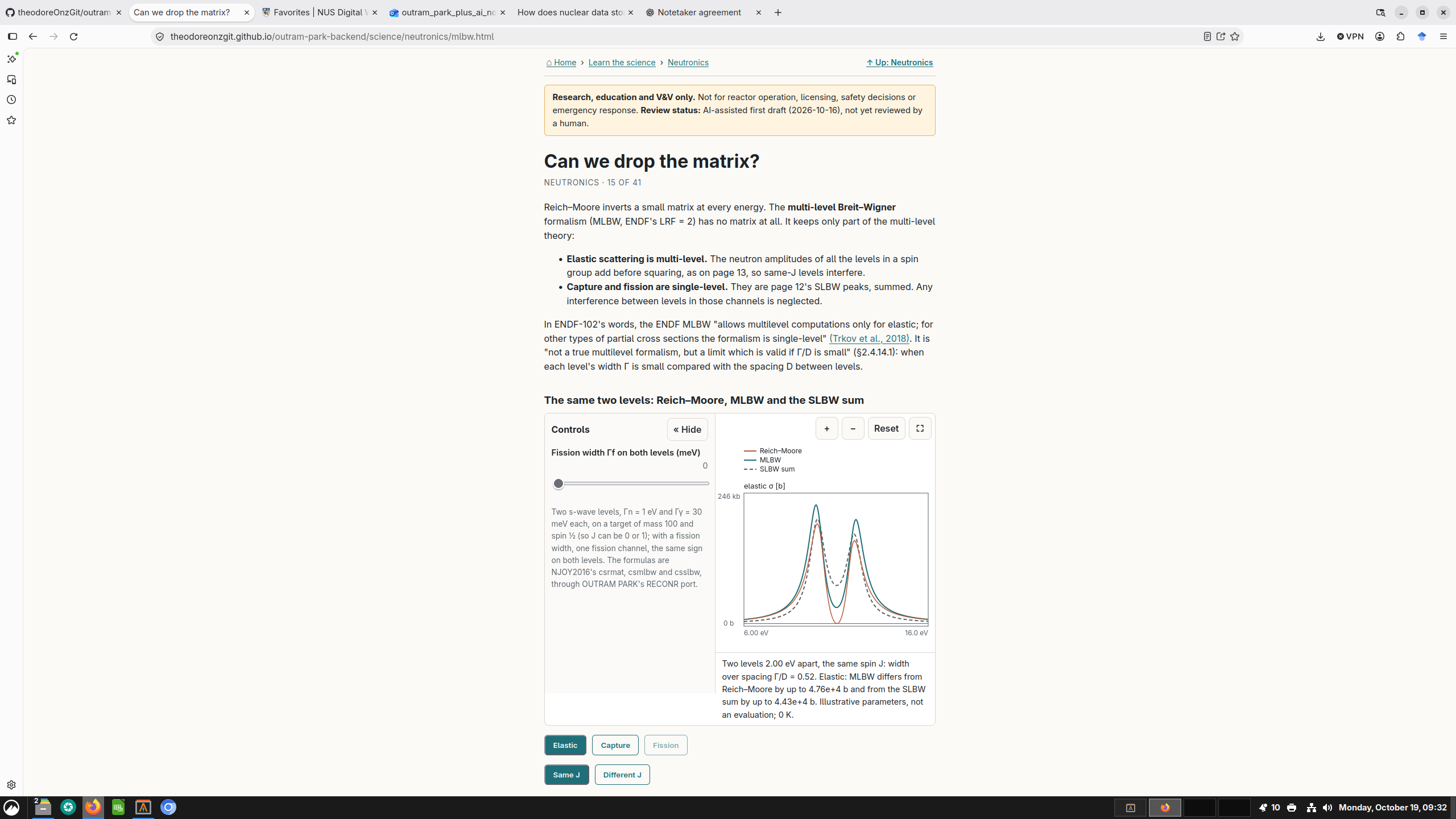}
\caption{Learn the Science page ``Can we drop the matrix?'', comparing the Reich--Moore, multi-level Breit--Wigner (MLBW) and single-level Breit--Wigner (SLBW) resonance formalisms. Page generated agentically (Claude Code) under the first author's direction; screenshot by the first author, 19 October 2026.}
\label{fig:docs_mlbw}
\end{figure}

Thus, in constructing the documentation agentically, the code was better documented, which helps new users understand jargon in NJOY manuals by relating them to physics concepts. Moreover, the first author also learned more about the formalisms after going through several iterations of documentation, helping him understand resonance formalisms with Opus and ChatGPT, thus preparing him to review code that was agentically coded.

This process was then repeated for the LEAPR documentation, where in the process of constructing LEAPR documentation agentically, the first author learned more about phonon interactions with neutrons and how individual phonon interactions summed up statistically, taking into account phonon vibrational modes form the basis for density of states. This then formed the baseline for thermal scattering laws used in lattice-based scattering of neutrons.

The documentation, at the time of publication, is still under review and construction. However, the process of agentic documentation with this human review not only generates the documentation, but also allows would-be documenters to learn about the code themselves. This not only prepares them for reviewing code, but also prepares human-reviewed AI-generated documentation which has been used for learning. Such a process is intended to produce high quality documentation from the outset of code generation, wherein agentic documentation with a human in the loop demonstrates an interactive improvement process through human feedback and use.

\section{Discussion}

Agentic V\&V provides a basis where a developer with limited expertise can translate two entire codebases accurately such that they reproduce the results for code-to-code verification with NJOY and OpenMC. Moreover, the acceptance testing has demonstrated the usability of the OutramMC web UI across Windows, macOS, Linux, iOS and Android devices. This increases the user friendliness of running these Monte Carlo code cases, even running benchmark calculations of Godiva and LCT-008 at high statistics (1500 active cycles, 50000 neutrons per cycle after demonstrated Shannon entropy convergence) on these devices. These serve as both qualification and acceptance testing for the web deployment of OutramMC and the Rust fork of NJOY.

Agentic V\&V was complemented by agentic document generation where web demos served not only as acceptance and qualification testing, but also as living documentation artefacts which help the user get interested in Outram Park, study the science behind Outram Park simulations and prepare the human agentic manager to review the code itself.

The result of this preparation process is a working web demo, verified code, with AI generated and human reviewed documentation that has gone through multiple rounds of iteration before reaching the end user. In this case, OutramMC and the Rust fork of NJOY have demonstrated code-to-code verification through the five routes, and validation with respect to Godiva, HST009 and LCT-008. This is a valuable contribution to the modernisation of legacy nuclear data and neutronics code as ACE files identical to those produced in NJOY were produced in the course of testing NJOY. Moreover, these serve as the first high-statistics validation runs done in a web browser over various platforms for these ICSBEP cases (to the best of the author's knowledge). With such modernisation, it is now possible that a browser version of NJOY which takes in ENDF files and prepares ACE files at any temperature can be done without a single line of code entered into the terminal. Moreover, any end user or academic could easily download the codes, or run the validation cases from the web browser to ascertain if the V\&V claims of this paper were correct. This enables a high degree of traceability and reproducibility.

Of course, reproducibility via web browser does not guarantee that the simulations are correct, but if authors wish to establish and prove that code is verified and validated, this claim would be easily testable through a web browser. This is much more user friendly as compared to compiling code and running the test cases.

\section{Threats to Validity}

This work applies mostly to nuclear-science-related codes. Some codebases do not have the rigour level required for nuclear science and research codes such as OutramMC and NJOY. For example, human review may not need to be performed for the codebases of a simple website. However, human review may be more important for the user interface and user experience (UI/UX) as opposed to the codebase itself, unless, of course, performance issues arise due to said faults of the codebase which cannot be agentically debugged.

Of course, for cases where websites have sensitive data or perform critical infrastructure functions, human review may still be required to complement AI review.

Additionally, this entire workflow depends on there being an upstream oracle, in this case the NJOY and OpenMC GitHub repositories. (Credit is due to the authors for designing and implementing the codebases and making them freely available). For cases in which there is no open-source code available, agents may have to scan literature in order to implement codes from scratch in a clean room implementation. Therein, a different approach needs to be taken. Moreover, this work only covers four ICSBEP papers, the full core HTGR double heterogenity calculations remain outside the scope of this work.

One particular limitation here is that while the first author is not an expert at coding NJOY or Monte Carlo algorithms, he did have training in nuclear engineering via classes in graduate school. This basic knowledge was still crucial in agentically preparing the documentation and web demos. Therefore, it cannot be claimed that a person with no expertise can translate any scientific codebase successfully using this methodology. However, coding agents lower the bar at which the methodology can be applied.

Lastly, the documentation process has not yet been tested against independent reviewers. This documentation has only been reviewed by the first author at the point of publication.

\section{Conclusion}

In this work, we have demonstrated the methodology of using agentic V\&V as well as agentic documentation via tutorials and web demos as a means of preparing a scientific codebase for human review. This methodology was applied to Rust translations and forks of OpenMC and NJOY. The byproduct of this is an agentically generated codebase with human-reviewed documentation which prepares the code developer to review code. This was demonstrated as Rust forks and translations of NJOY and OpenMC were able to reproduce behaviour by producing several ACE files necessary for Monte Carlo code, and have code-to-code verification and validation run on four ICSBEP benchmarks. Moreover, preparing code in such a manner prepared the codebase for human review. This human review process is especially important for nuclear science codes for research wherein high quality standards are expected, and cannot be neglected even when codes are agentically generated.

The codes produced also demonstrated deployment to web browsers across five platforms: Android, iOS, Windows, macOS and Linux. The acceptance tests served as benchmarking tests as well when Monte Carlo runs were done at high statistics. This enabled runs to be done fairly quickly, even when servers or desktops were not available, as certain Monte Carlo runs could simply be done on webpages. This end-to-end nuclear data processing to criticality benchmark in a web browser is a novel approach (to the best of the author's knowledge) to running criticality calculations and benchmarks, and could be a convenient means with which to run scientific research codes in future.

\subsection{Future Work}

Human code review is the next step after agentic V\&V and documentation, corresponding to the third rung of maturity in Outram Park. This will be attempted in future work with agentically generated Outram Park code with respect to codes relevant to high-temperature gas-cooled reactor neutronics and source term calculations.

The author also welcomes nuclear experts who are proficient in nuclear data or neutronics to participate in future human-based reviews of NJOY and OpenMC algorithms as well, as in the NRC documentation, it is recommended that qualified personnel should be allowed to be a second reviewer for code generated with humans. \cite{nureg_br_0167}

\section*{Acknowledgements}
\addcontentsline{toc}{section}{Acknowledgements}

The author would like to thank Associate Professor Chung Keng Yeow, Director of the Singapore Nuclear Research and Safety Institute (SNRSI), and Low Xin Wei, CEO of SNRSI, for their support in developing Outram Park for SNRSI.

The author also thanks colleagues Vitesh and Zi Liang for running the Godiva and LCT-008 web demonstrations on Windows as part of UI/UX testing, which was in turn important for acceptance testing. Their testing helped identify limitations in the web user interface, including restrictions on the number of source neutrons per cycle and the number of simulation cycles, as well as the lack of multithreaded execution for LCT-008 in the browser.

The author would also like to thank his cousin, Janette Loh, for conducting the HST-009 acceptance and validation tests on her iPad Air, and his girlfriend, Wong Jing Ting, for conducting the LCT-008 acceptance and validation tests on her MacBook Air.

This work was performed with funding support from the Singapore Nuclear Research and Safety Institute (SNRSI). NRF funding number (A-8002968-00-00).

\section*{CRediT Authorship Contribution Statement}
\label{sec:credit}

\textbf{Theodore Kay Chen Ong:} Conceptualization; Methodology; Software;
Validation; Formal analysis; Investigation; Data curation; Visualization;
Writing --- original draft; Writing --- review \& editing; Project
administration.
\textbf{Sicong Xiao:} Supervision; Writing --- review \& editing.

\vskip 2em

\section*{Declaration of generative AI and AI-assisted technologies in the manuscript preparation process}
\addcontentsline{toc}{section}{Declaration of generative AI and AI-assisted technologies in the manuscript preparation process}

Generative artificial intelligence (AI) tools, including large language models and agentic coding systems, were extensively used in this work. Anthropic Claude models, including Opus 4.8, 5.0 and 5.5, as well as Sonnet and Haiku 5.0 and 5.5, were used during different stages of development. Multiple AI subagents were employed, and an exhaustive record of individual model assignments was not maintained. OpenAI ChatGPT, including ChatGPT Sol accessed through the National University of Singapore (NUS) ChatGPT Edu service, was also used. Microsoft Enterprise Copilot and NUS AI Know, the National University of Singapore's institutional AI assistant (which routes prompts to Anthropic Claude, OpenAI ChatGPT and Google Gemini backends), were used for initial ideation and discussion. They were not suitable for agentic coding: at the time of use they operated as conversational assistants only, without the ability to autonomously read, edit, run and iterate on a multi-file Rust workspace.

AI coding agents were employed to translate, implement, test, debug, document and deploy the scientific software described in this paper, including the Rust port of NJOY and OutramMC, a Rust implementation derived from OpenMC. AI agents also generated unit and integration tests, performed differential comparisons against upstream software, investigated numerical discrepancies, prepared benchmark calculations and produced summaries of verification and validation results.

Generative AI tools were additionally used for literature discovery, scientific explanations, technical discussions, documentation development and manuscript editing. The scientific arguments, research objectives, methodology, interpretation of results and conclusions were directed by the human author. AI-generated documentation underwent iterative human review and revision, as described in the manuscript.

The author acknowledges that agreement with upstream software and experimental benchmarks does not establish the correctness of every underlying algorithm. The software and documentation remain subject to further human review, including review by appropriately qualified domain experts.

The author retains responsibility for the scientific claims, interpretation of results, accuracy of references and final content of this manuscript. AI tools are not recognised as authors of this work.

\bibliographystyle{unsrt}
\bibliography{./outram_park_2026}

\end{document}